\documentclass{jfm}  
\usepackage{natbib}
\usepackage{graphicx}
\usepackage{epstopdf, epsfig}
\usepackage{graphicx,subfig}
\usepackage{amsmath}
\usepackage{color}
\usepackage{calligra}

\DeclareMathAlphabet{\mathcalligra}{T1}{calligra}{m}{n}

\newcommand{\Fourier}[1]{\mathcal{F}\left(#1\right)}

\newcommand{\ii}{\mathrm{i}}
\newcommand{\e}{\mathrm{e}}

\newcommand{\vv}[1]{{\boldsymbol{#1}}}
\newcommand{\mm}[1]{\mathsfbi{#1}}

\newcommand{\x}{\vv{u}}
\newcommand{\f}{\vv{f}}
\newcommand{\y}{\vv{y}}
\newcommand{\w}{\vv{w}}
\newcommand{\z}{\vv{z}}
\newcommand{\n}{\vv{n}}

\newcommand{\A}{\mm{A}}
\newcommand{\B}{\mm{B}}
\newcommand{\C}{\mm{C}}

\newcommand{\R}{\mm{R}}
\newcommand{\Ry}{\mm{R}_{\y}}
\newcommand{\I}{\mm{I}}
\newcommand{\F}{\mm{F}}

\newcommand{\U}{\mm{U}}

\newcommand{\V}{\mm{V}}

\newcommand{\W}{\mm{W}}

\author{Eduardo~Martini\aff{1,2}
	\corresp{\email{eduardo.martini@univ-poitiers.fr}}, 
	Daniel~Rodr\' iguez\aff{3}, Aaron~Towne\aff{4},
	Andr\' e~V.~G.~Cavalieri\aff{1}}
\shortauthor{E.~Martini, D. Rodr\' iguez, A.~Towne, A.~Cavalieri}

\affiliation{\aff{1}Instituto Tecnol\' ogico de Aeron\'{a}utica,   S\~ ao Jos\' e dos Campos/SP - Brazil \\ 
	\aff{2} D\'epartement Fluides, Thermique et Combustion, Institut Pprime, CNRS, Universit\' e de Poitiers,ENSMA, 86000 Poitiers, France 
	\aff{3}{ETSIAE-UPM (School of Aeronautics) - Universidad Polit\'ecnica de Madrid, Spain.} \\
	\aff{4} Department of Mechanical Engineering, University of Michigan, Ann Arbor, MI 48109, USA}

\title{Efficient computation of global resolvent modes}
\shorttitle{Efficient computation of global resolvent modes}

\begin{document}
	\maketitle
	\begin{abstract}
		
		Resolvent analysis of the linearized Navier-Stokes equations provides useful insight into the dynamics of transitional and turbulent flows and can provide a model for the dominant coherent structures within the flow, particularly for flows with large gain separation.  Direct computation of force and response modes using a singular value decomposition of the full resolvent matrix is  feasible only for simple  problems; despite recent progress, the cost of resolvent analysis for complex flows remains considerable. In this paper, we propose a new matrix-free method for computing resolvent modes based on integration of the linearized equations and the corresponding adjoint system in the time domain.  Our approach achieves an order of magnitude speedup when compared to previous matrix-free time stepping methods by enabling all frequencies of interest to be computed simultaneously. Two different methods are presented: one based on analysis of the transient response, providing leading modes with fine frequency discretization; and another based on the steady-state response to a periodic forcing, providing optimal and suboptimal modes for a discrete set of frequencies. 
		The methods are validated using a linearized Ginzburg-Landau equation and applied to the three dimensional flow around a parabolic body.   
	\end{abstract}
	
\section{Introduction}

	
	Resolvent analysis constitutes an input-output framework between forces and their responses in the frequency domain.  This approach has attracted the attention of the fluid mechanics community after \citet{mckeon2010criticallayer} used it to model a turbulent channel flow, showing that if forcing terms show no preferential direction the flow response is dominated  by the optimal response mode.   In this case, \cite{towne2018spectral} showed that these optimal response modes provide an approximation of coherent structures within the flow as defined by spectral proper orthogonal decomposition.  
	 Several studies applied the same ideas to other flows \citep{beneddine2016conditions,abreu2017coherent,lesshafft2019resolvent,schmidt2018spectral,yeh2019resolvent,abreu2020wavepackets}, and to develop estimation methods \citep{gomez2016reduced,sasaki2017realtime,beneddine2017unsteady,symon2018reconstruction,towne2020resolvent,martini2020resolventbased}. 
	
	If the flow has one non-homogeneous direction, resolvent modes and gains can be obtained by direct manipulation of the matrix that represents the discretized system \citep{mckeon2010criticallayer}.  When a direct matrix decomposition is not possible, iterative methods are needed. These can typically have two  parts: (a) obtaining the effect of the resolvent operator acting on a vector, and (b) algorithms that use (a) to approximate singular values and vectors. To distinguish these, we will refer to (a) as ``methods", and to (b) as ``algorithms".
	
	Different methods have been used in the literature. The effect of the resolvent on a vector can be obtained by solving a linear problem. If the matrix that describes the system can be constructed, a LU factorization can be used to solve the linear system and obtain resolvent modes iteratively \citep{schmidt2018spectral,ribeiro2020randomized}.   \cite{brynjell-rahkola2017computing} solved the linear problem using  a GMRES method, which was accelerated with the use of pre-conditioners on flows with low Reynolds numbers. \citet{monokrousos2010global} used a matrix-free approach, using time marching of the direct and adjoint equations. On each iteration the system was harmonically forced with the previous iteration result until the steady-state-response was reached, repeating the method until convergence provides optimal force and response modes for a given frequency.
	
	Power-iteration algorithms are popular \citep{monokrousos2010global}. More advanced algorithms, using Krylov spaces and Arnoldi methods have been mentioned in the literature \citep{monokrousos2010global}, but to the best of the authors knowledge not used with time stepping methods in previous works.  Recently randomized singular-value decompositions (rSVD) were proposed by \cite{ribeiro2020randomized}, showing  an algebraic convergence rate with the number of random vectors. 
	
	Alternatively, reduced-order models (ROMs) have been used to approach such systems, e.g. ROMS based on the system eigen-modes  \citep{akervik2008global,alizard2009sensitivity,schmid2012stability}.  However such a basis can be a bad choice to describe the system \citep{trefethen1997pseudospectra,rodriguez2011towards,lesshafft2018artificial}, and it is not necessarily clear what is a proper choice of basis for a given system. 
	
	In this study we propose two matrix-free methods, one being an improvement of the method proposed by \citeauthor{monokrousos2010global}, and another an adaptation of methods used on a previous study \citep{martini2020resolventbased}, to compute resolvent gains and modes for several frequencies simultaneously. The solutions of the system's frequency-domain representation are obtained for several frequencies simultaneously via time marching of the system's linearized equations.  The simultaneous solution of resolvent modes for multiple frequencies provided by these two methods allows for a substantial reduction of the overall computational effort.
	
	The paper is organized as follows: \S~\ref{sec:basicEq} provides the basic equations for resolvent analysis. The proposed methods are presented in \S~\ref{sec:transResponse} and \S~\ref{sec:SSResponse}, with a discussion of their costs and best practices  in \S~\ref{sec:discussion}.  An application on a Ginzburg-Landau problem, illustrating expected trends, is presented in \S~\ref{sec:GL}. Resolvent analysis on the flow around a parabolic body is presented \S~\ref{sec:parabolicBody}, and final conclusions in \S~\ref{sec:conclusions}.	
	
	\section{Frequency-domain iterations using time marching} 
	\subsection{Basic equations} \label{sec:basicEq}
	We work with a stable linear system given, in discretized form, by
	\begin{align}\label{eq:system}
	\begin{aligned}
	\dfrac{d }{d t}\x(t) + \A \x(t) =&  \B\f(t), \\
	\y(t) =& \C \x(t),
	\end{aligned}
	\end{align}
	where  $ \x $, $ \f$ and $ \y $ are columns vectors representing the system state, driving force and observations, with sizes $ n_u$ , $n_f $ and $ n_y $, respectively. The matrix $ \A$ ($ n_u\times n_u $) defines the system  dynamics, i.e. the linearized Navier-Stokes equations. The matrices $ \B $ ($ n_u\times n_f $) and  $ \C $ ($ n_y\times n_u $) correspond to forcing and observation matrices, respectively.
	
	The solution of such a system can be obtained as a combination of the inhomogeneous solution, a given $ \x(t) $ that satisfies \eqref{eq:system}, to which a linear combination of homogeneous solutions is added in order to satisfy a prescribed initial condition. The inhomogeneous solution can be expressed in the frequency domain as
	\begin{align}
	\hat \x(\omega) =& \;\;\;\R(\omega) \B \hat \f(\omega) ,\\
	\hat \y(\omega) =& \C \R(\omega) \B \hat \f(\omega) = \Ry \hat \f(\omega),
	\label{eq:systemFreqDomain}
	\end{align}
	where hats denote the Fourier transform,
	\begin{align}\label{eq:fourier}
	\hat {(\cdot)} = \Fourier{{(\cdot)} }= &\int_{-\infty}^{+\infty}  (\cdot) \mathrm{e}^{\ii\omega t} dt.
	\end{align}
	 The resolvent operator is defined as  $ \R(\omega) =  \left(-\ii\omega \mm I +\A\right)^{-1}  $ and  $ \Ry=\C\R\B $. 
	
	Resolvent analysis consists finding optimal force components, which maximize gains defined as
	\begin{align}
		G(\omega) = \frac {\lvert\lvert \hat \y(\omega) \rvert}{\lvert\lvert \hat \f(\omega)\rvert\rvert }=\frac {\lvert\lvert\Ry\hat \f(\omega) \rvert\rvert}{\lvert\lvert\hat \f(\omega)\rvert\rvert}. 
	\end{align}
	Such gains and modes can be obtained via a singular-value decomposition (SVD) of $ \Ry $, which reads $ \Ry=\U \Sigma \V^\dagger $,  where $ \U $ and $ \V $ are unitary matrices containing response ($ \vv{ \mathcal  U}_i $) and force ($ \vv{\mathcal V}_i $) modes on their columns, and $ \Sigma $ is a diagonal matrix containing the non-negative singular values $ \sigma_i $, with $ {\sigma_1\ge\sigma_2\ge... \ge\sigma_n}$.   Due to their physical interpretation, left and right singular vectors will be respectively referred to as \emph{response} and \emph{forcing modes}, and singular values refereed as \emph{gains}.  \cite{mckeon2010criticallayer} used $ \B=\I $ and $ \C=\I $, with the physical interpretation that forces and responses anywhere in the flow have the same weight. Using different $ \B $ and $ \C $ matrices allows for localization and weighting of forces and responses in space. 
	
	The adjoint equations corresponding to \eqref{eq:system} are
	\begin{align}\label{eq:adsystem}
	\begin{aligned}
	-\dfrac{d }{d t}\z(t) + \A^\dagger \z(t) =&  \C^\dagger \y(t), \\
	 \w(t) =& \B^\dagger \z(t),
	\end{aligned}
	\end{align}
	were $``  \dagger "$ represents the adjoint operator for a suitable metric.
	 As non-uniform meshes are typically necessary in studies of complex flows, we assume generic metrics on the response~($ \mm W_u $), force~($ \mm W_f $) and observation~($ \mm W_y $) spaces, such that
	\begin{align}
		\langle \x_1 , \x_2 \rangle_u = \x_1^H \mm W_u \x_ 2,
	\end{align}
	where $``H"$ denotes the Hermitian transpose. Similar expression for force and observations spaces are used. The discrete adjoints are given by
	\begin{align}
		\A^\dagger =& \mm W_u^{-1} \A^H \mm W_u,  & 
		\B^\dagger =& \mm W_f^{-1} \B^H \mm W_u,  &
		\C^\dagger =& \mm W_y^{-1} \C^H \mm W_u.  
	\end{align}
	
	The corresponding frequency domain equation is given by
	\begin{align}
		\hat \z(\omega) =& \;\;\;\;\,\R^\dagger(\omega) \C^\dagger \hat \y(\omega) , &
		\hat \w(\omega) =& \B^\dagger \R^\dagger(\omega) \C^\dagger \hat \y(\omega) = \Ry^\dagger \y(\omega).
		\label{eq:adsystemFreqDomain}
	\end{align}

	For a given system reading component ($ \hat \y $), the adjoint equations provides the sensitivity ($ \hat \w $) of this reading to applied forces. 
	Note that $ \Ry^\dagger \Ry  = \V \Sigma^2 \V^\dagger$, and thus singular values and right singular vectors of $ \Ry $ can be obtained from an eigen-value decomposition of $ \Ry^\dagger \Ry $, i.e. using the adjoint problem a singular value problem can be converted into an eigenvalue problem.

	An explicit construction of $ \Ry$ require the storage and inversion of matrices, and can be unfeasible for large systems. Instead, matrix-free methods to obtain the results of $ \Ry $ applied to a given vector are used. To obtain such results for several frequencies simultaneously, the relation between the time and frequency domains is explored. 
	
	From a given  $ \f(t) $, its corresponding response can be computed using \eqref{eq:system}. 
	As the system is stable, and using as initial condition $ \x(t\to-\infty)=0 $, \eqref{eq:systemFreqDomain} provides the full solution to \eqref{eq:system}, as all homogeneous solutions diverge for $ t\to -\infty $. 
	Likewise, solutions for the adjoint problem, \eqref{eq:adsystemFreqDomain}  can be obtained from \eqref{eq:adsystem} when the terminal conditions $ {\f(t\to\infty)=0} $ is used. 	
	
	In practice, these solution can be obtained by time marching: if $ \f(t) $ is zero, or negligible, for $ t<t_0 $, $ \x(t) $ can be obtained by integration \eqref{eq:system} starting from $ t_0 $ using $ \x(t_0)=0$.  The solution for \eqref{eq:adsystem} can be similarly obtained if $ y(t)=0 $ for $ t>t_0 $ via an integration backwards in time. Fourier components of $ \x,\y,\f$ and $\w $ can be obtained with Fourier transforms of their corresponding time signals.
	
\begin{figure}
		\centering
		\includegraphics[width=\linewidth]{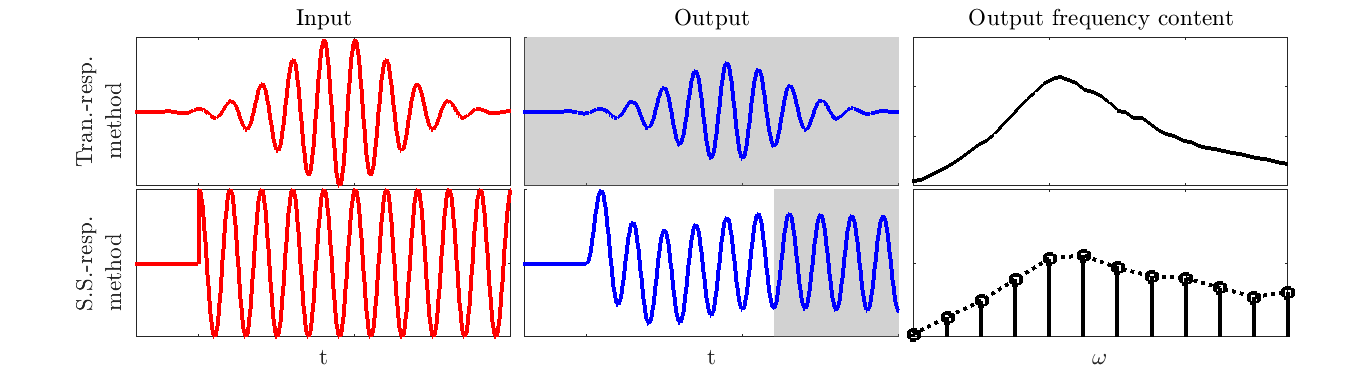}
		\caption{Illustration of the TRM (top) and SSRM (bottom). The shaded  area corresponds to the interval used to estimate Fourier coefficients. Input and output of the direct(adjoing) run are forces(readings) and readings(sensitivities). }
		\label{fig:methodsillustration}
	\end{figure}

	Time marching of \eqref{eq:system} and \eqref{eq:adsystem} will be referred to as the direct and adjoints run, respectively. Using readings $\y$ of the direct run, noting that  from \eqref{eq:systemFreqDomain}, $\hat \y = \Ry \hat \f$, as forcing terms of the adjoint run,   \eqref{eq:adsystemFreqDomain}  gives  $\hat \w = \Ry^\dagger \hat \y =\Ry^\dagger \Ry \hat \f $, i.e. the action of $\Ry^\dagger \Ry $ on a given vector is computed from  the direct and adjoint runs. This computation allows the use of the algorithms presented in Appendix~\ref{app:algorithms} to compute eigenvalues of $ (\Ry^\dagger \Ry)$, and thus singular values of $ \Ry$.  Two  methods to compute the action of $ (\Ry^\dagger \Ry) $ on a vector are presented next: the transient-response method (TRM) , detailed in \S \ref{sec:transResponse}; and the steady-state-response method  (SSRM), detailed in \S \ref{sec:SSResponse}. An illustration of these methods is shown in figure \ref{fig:methodsillustration}.

	\subsection{Transient-response method (TRM)} \label{sec:transResponse}
	This method uses the full response obtained from time marching \eqref{eq:system} and \eqref{eq:adsystem} to compute solutions of \eqref{eq:systemFreqDomain} and \eqref{eq:adsystemFreqDomain}.  Using a compact force in time force on \eqref{eq:system}, a compact response, in the sense that it is exponentially decaying for large times, is obtained.  Such responses, when used as external forces in \eqref{eq:adsystem} also lead to compact responses. Taking the Fourier transform of the external forces used on  \eqref{eq:system} and responses of  \eqref{eq:adsystem} provides $\hat \w = \Ry^\dagger \Ry \hat \f $. The approach is illustrated in figure \ref{fig:methodsillustration}.
	
	Assuming compactness of the forcing terms of  \eqref{eq:system}, the stability of the system guarantees that the Fourier transforms of $ \x,\y,\f$ and $\w $ are always well defined. This follows from the fact that  $ \x(t) \approx \e^{-\omega_i t} $ for large $ t $, where $ \omega_i $ is the imaginary part of the least-stable mode. As $ \x(-\infty)=0 $, $ \Fourier{\x(t)} $ is well defined,  so is $ \Fourier{\y(t)} $. A similar argument holds for the adjoint system. 
	
	For finite-precision numerical computations, it is important that the frequency content of $ \f(t) $ is normalized, ensuring that signals from frequencies with larger gains do not contaminate other frequencies due to finite precision and sampling: this will be illustrated in \S~\ref{sec:GL}. Normalization can be performed using a temporal filter which flattens the energy's power-spectral density (PSD) over the desired frequency range. In this work finite impulse response (FIR) filters are used \citep{press2007numerical}. FIR filters guarantee that the exponential decay present in the signals described above is maintained. An overview on this class of filters and trends obtained for spectra flattening are presented in appendix \ref{app:filters}. 
 
	In practice, integration and filtering only needs to be performed in time until the energy norm of the flows becomes negligible, after which the time series can be truncated. Spectral leakage, which is an expected consequence of truncating the time domain, is proportional to the signal's value at its edge. As the signals here show an exponential decay for large $ |t| $, such error decreases exponentially with the total integration time. 
	
	Using the power-iteration algorithm, described in appendix~\ref{app:algorithms}, reading from the adjoint run, $\w$, are used as forcing terms of a new pair of direct and adjoint runs.  Iterating the procedure $ \hat \y$ and $\hat \w$ converge to the leading response and force modes.
	In order to use results of a direct run into a adjoint run, and vice-versa, checkpoints of the simulation need to be saved to disk. During the subsequent run, the solution of the previous run needs need to be loaded and interpolated to construct the forcing term for the present equation.  Different interpolations methods are presented and compared in appendix \ref{app:interpSpecProp}. Note that sampling frequency and filter order are linearly related for a constant filter frequency resolution. To simultaneously reduce storage size, interpolation errors and filter order, the $ C^2 $ interpolation is recommended.
 
	\subsection{Steady-state-response method (SSRM)} \label{sec:SSResponse}
	In contrast with the TRM, where the solutions of the direct and adjoint equations to excitations localized in time are used, the stead-state method is based on the system's steady-state-response to periodic excitations. An initial periodic force with period $T$ is constructed as
	\begin{equation}\label{eq:forceFseries}
		\f(t) = \begin{cases} 
		\Real ( \hat \f (\omega_0) )  +  2 \sum_{k=1}^{n_f} \Real \left( \hat \f(\omega_k) \e^{-\ii\omega_k t} \right) & \text{,for real $\f$} \\
		   \sum_{k=0}^{n_f}  \hat \f(\omega_k) \e^{-\ii\omega_k t}  & \text{,for complex $\f$} 
		\end{cases},
	\end{equation}	
	where $ \omega_k = 2\pi k /T $, corresponding to a Fourier series with $ n_f $ coefficients.  Fourier series coefficients for $ \hat \y(\omega_k) $ are obtained via the steady-state, time-periodic response of \eqref{eq:system}	and used to construct an excitation term for \eqref{eq:adsystem} with an expression similar to \eqref{eq:forceFseries}. 	The terms $ \f(t) $ and $ \w(t) $ are the iteration input and output.
	Combining the steady-state response of \eqref{eq:system} and \eqref{eq:adsystem}, the action of $ \Ry^\dagger\Ry $ on $\hat \f $ is obtained.
	
	The time scale at which the transient responses vanishes, and thus the state converges to the steady-state response, can be estimated from a prior run where the norm of random initial condition reaches a prescribed small value. Fourier-series coefficients for forces and responses are obtained via Fourier transforming a time block of length $ T $ after transients have vanished, as illustrated in figure \ref{fig:methodsillustration}.

	The SSRM provides the action of $\Ry^\dagger\Ry$ on a given vector, which can be used for computation of gains and modes for a discrete set of frequencies, with frequency resolution given by $ \Delta \omega = 2\pi / T $, using the algorithms presented in appendix~\ref{app:algorithms}. The normalisation  of amplitudes,  needed for the power-iteration algorithm, and orthogonalization of inputs, needed for the Arnoldi algorithm, can be performed  on the Fourier series components, avoiding the need for a temporal filter and of saving checkpoints.   
	
	\subsection{Discussion}\label{sec:discussion}
	 TRM and SSRM have different characteristics that make them suitable for different applications. These differences and guidelines for the choice of method and algorithm are presented here. We focus on two classical algorithms:  power-iteration and Arnoldi. These are briefly reviewed in appendix \ref{app:algorithms}.
	
	The main parameter in the TRM is the sampling rate, which needs to be defined in terms of the cut-off frequency. Below the cut-off frequency, all frequencies can be resolved simultaneously, which can be obtained either by zero padding the time series prior to using an FFT algorithm, or by using \eqref{eq:fourier} directly. The approach is thus better suited if fine frequency discretization is desired.  While frequency normalization of inputs can be obtained using only one filter application, their orthogonalization requires time series filtering and additions, and the repeated application of these steps required by the Arnoldi algorithm becomes costly and prone to error accumulation for large systems.  Thus, the TRM is better suited for the power iteration algorithm, which limits its applicability to problems in which only the leading resolvent mode is of interest.  
	
	
	The main parameter of the SSRM is the time length used to characterise the steady-state response. This length is given by the periodicity of the signal, $ 2\pi/\Delta \omega $, where $ \Delta \omega $ is the desired frequency discretization. This method is better suited if coarser frequency discretization can be used, and in particular if one is interested in higher frequencies, which add little to the computational cost. It is straightforward to use either the power-iteration or Arnoldi algorithms with it. If suboptimals are desired, a combination of the SSRM with the Arnoldi algorithm should be used.
	
	Note that the SSRM has roughtly the same cost as the one proposed by \cite{monokrousos2010global}, but provides modes and gains for several frequencies simultaneously: if $ n_\omega $ frequencies are desired, then the SSRM is, approximately,  $ n_\omega $ times cheaper. Assuming $ n_\omega > 10$,  total costs can be reduced by more than an order of magnitude, making the method comparable to the preconditoned approach used by \cite{brynjell-rahkola2017computing}. Unlike their approach, the  method proposed here  is not limited to cases with low-Reynolds numbers. 
	
	Note also that the formulation derived here is general, and can be implemented on any solver of direct and adjoint linearized Navier-Stokes equations. Naturally, the total computational cost is directly dependent on the solver performance. It is beyond the scope of this work to compare different DNS
	 approaches, but is worth mentioning that efficient codes and methods are currently available that show good scalability with the number of degrees-of-freedom and parallelization.

	\section{Validation and trends for the Ginzburg-Landau equation}
	\label{sec:GL}
		We explore the properties of the method using a linearized Ginzburg-Landau (GL) model, for which resolvent gains and modes can be directly obtained by manipulation of the system matrices. The model qualitatively mimics the behavior of some complex flows, and has been widely used to explore tools and methods \citep{chomaz1991frequency,couairon1999fully,bagheri2009amr,cavalieri2019amr,martini2020resolventbased}. The model is given by
		\begin{align} 
		\frac{\partial \x(x,t)}{\partial t} +  \A  \x(x,t)  =& \f(x,t) \label{eq:GL_timeDomain}, &
		\A &= U \frac{\partial }{\partial x} - \mu(x) - \gamma \frac{\partial^2 }{\partial x^2},
		\end{align}
		and we here use parameters $ U=6 $, $ \gamma=1 $ and $ \mu(x)=\beta \mu_c(1-x/20) $, where $ \mu_c=U^2 \Real (\gamma) / |\gamma|^2$ is the critical value for onset of absolute instability \citep{bagheri2009amr}. The parameters are similar to those used by \cite{lesshafft2018artificial}, but here we choose to keep $ \gamma $, and therefore the equation and its solution, real.  The terms in $ \A  $ correspond to advection, growth/decay and diffusion, respectively. Dirichlet boundary conditions are considered at $ x=0 $ and $ 40 $, $ \x(0,t)=\x(40,t)=0 $, and the initial condition $ \x(x,0)=0 $ is used. We consider a system with $ \beta= 0.1$, leading to a moderate gain separation  between optimal and suboptimal modes, evaluated via singular-value decomposition of the resolvent operator. For simplicity, we assume $ \B=\C=\I $.
		
		Starting from an impulse-like in time and random in space force vector, $ \f(t) = \f_0 \delta(t) $, which was implemented as an initial condition, direct and adjoint runs were performed using a time step of $ 10^{-2} $ using a second-order Crank-Nicolson scheme. Gains and modes for  frequencies up to $ \omega = 15 $,  were accurately recovered. Time marching was carried out until the state norm is lower than $ 10^{-9} $.
		
\begin{figure}
			\centering
			\subfloat[Estimated gain]{\includegraphics[width=0.475\linewidth,trim={0 0 518 0},clip]{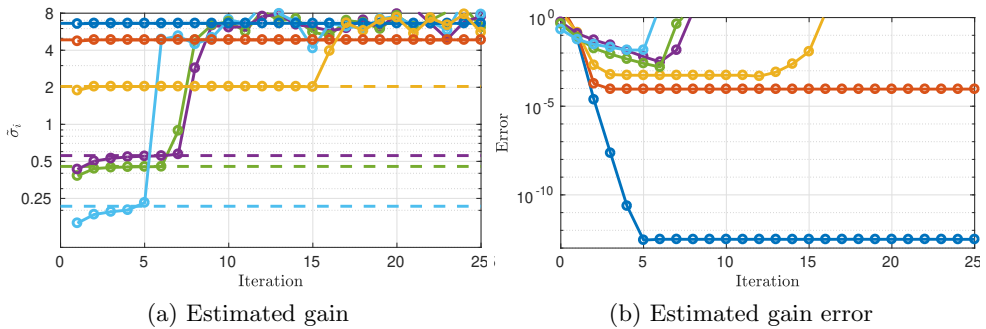} }
			\subfloat[Estimated gain error]{\includegraphics[width=0.475\linewidth,trim={500 0 10 0},clip]{Figures/Gains_notfiltered.eps} }
			
			\caption{Leading resolvent gains using the iteration scheme  without regularisation. (a) leading resolvent estimated gains ($\tilde \sigma$) for different frequencies as a function of iteration count: solid lines represent iteration gains and dashed line the true optimal gains. (b) gain error, $ |1-\tilde \sigma_1/\sigma_1| $.}
			\label{fig:covergencegainsnotfiltered}
		\end{figure}

\begin{figure}
			\centering
			\subfloat[Spectral amplitude]{\includegraphics[width=0.475\linewidth,trim={0 0 518 0},clip]{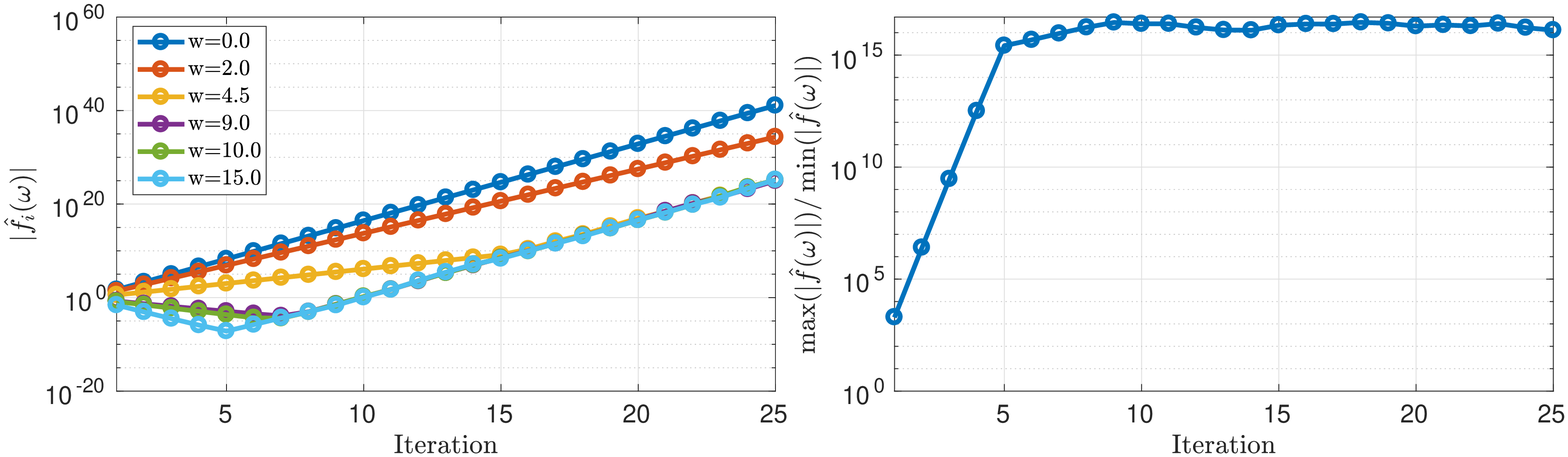} }
			\subfloat[Condition number]{\includegraphics[width=0.475\linewidth,trim={509 0 15 0},clip]{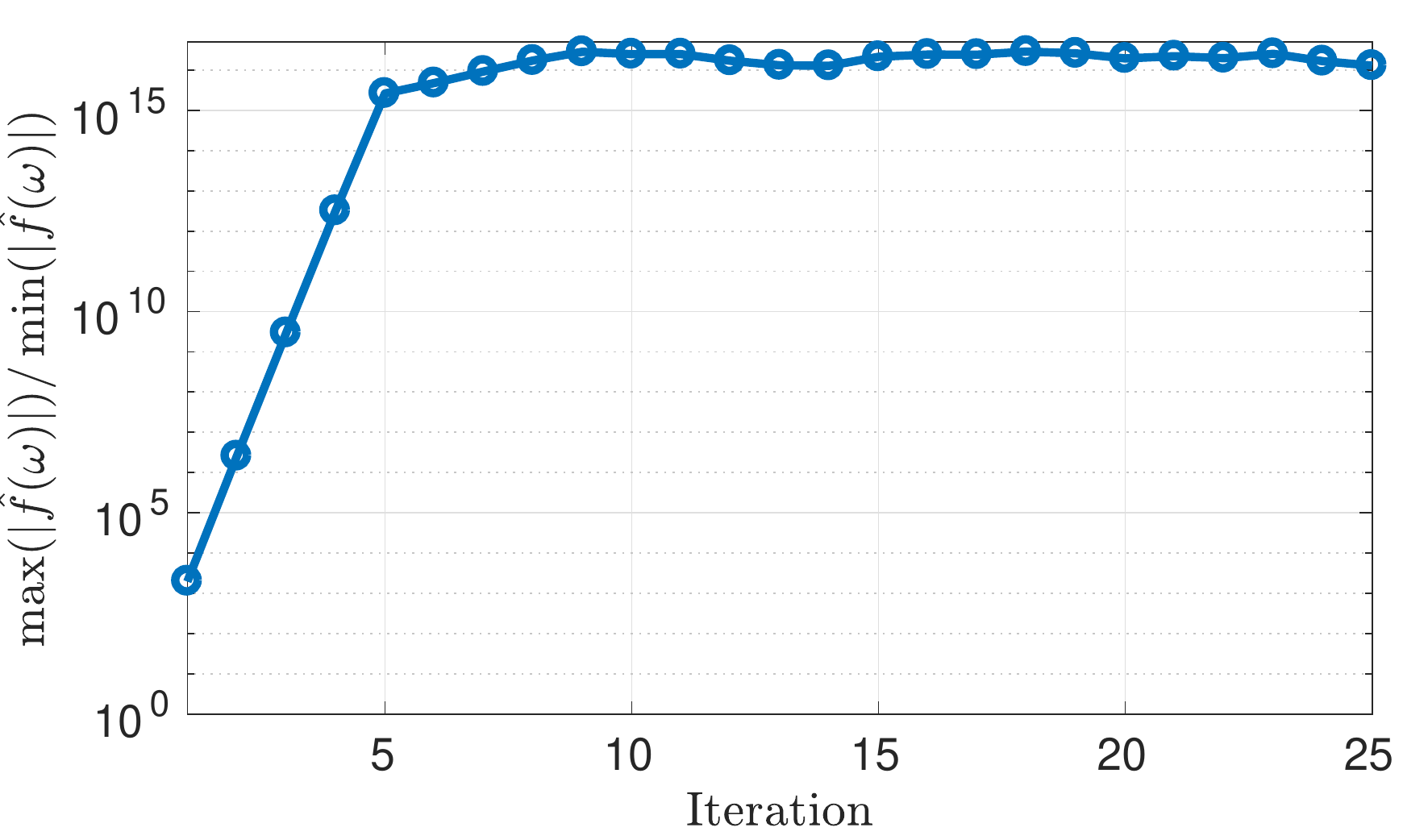} }
			
			\caption{(a) evolution of the amplitudes of different spectral components of the un-regularized iteration scheme. (b)  condition number given by the ratio between the largest and smallest spectral components.}
			\label{fig:amplitudesnotfiltered}
		\end{figure}

		Figure \ref{fig:covergencegainsnotfiltered} illustrates the evolution of gain estimation using the power-iteration algorithm when normalization is not performed. Gains for low frequencies converge to the true values, while for larger frequencies gains seem to approach them, but after further iteration diverge and oscillate around the maximum gain. Figure \ref{fig:amplitudesnotfiltered}(a) shows the evolution of the norm of each spectral component: it is apparent that each spectral component has its own amplification trend until the ratio between its amplitude and the largest amplitude becomes $ \approx 10^{-16} $. This is confirmed by the condition number shown in figure \ref{fig:amplitudesnotfiltered}(b), which saturates at $ 10^{16} $. At this point numerical errors from the larger components dominate the signal at these frequencies.

		
\begin{figure}
			\centering
			\subfloat[Spectral amplitude]{\includegraphics[width=0.475\linewidth,trim={0 0 515 0},clip]{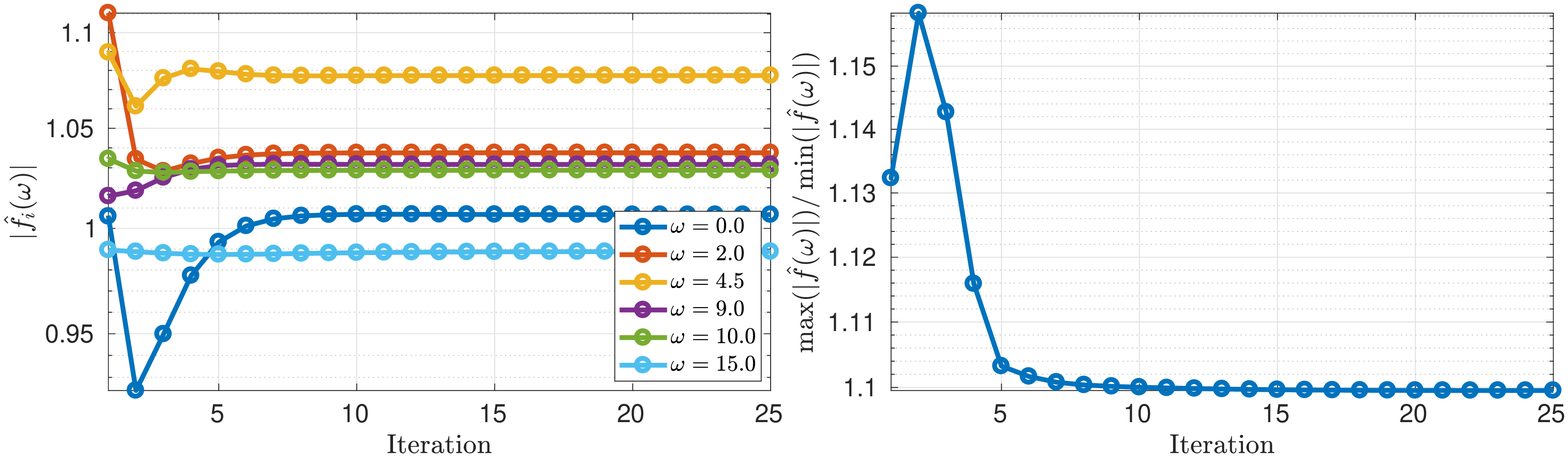} }
			\subfloat[Condition number]{\includegraphics[width=0.475\linewidth,trim={508 0 15 0},clip]{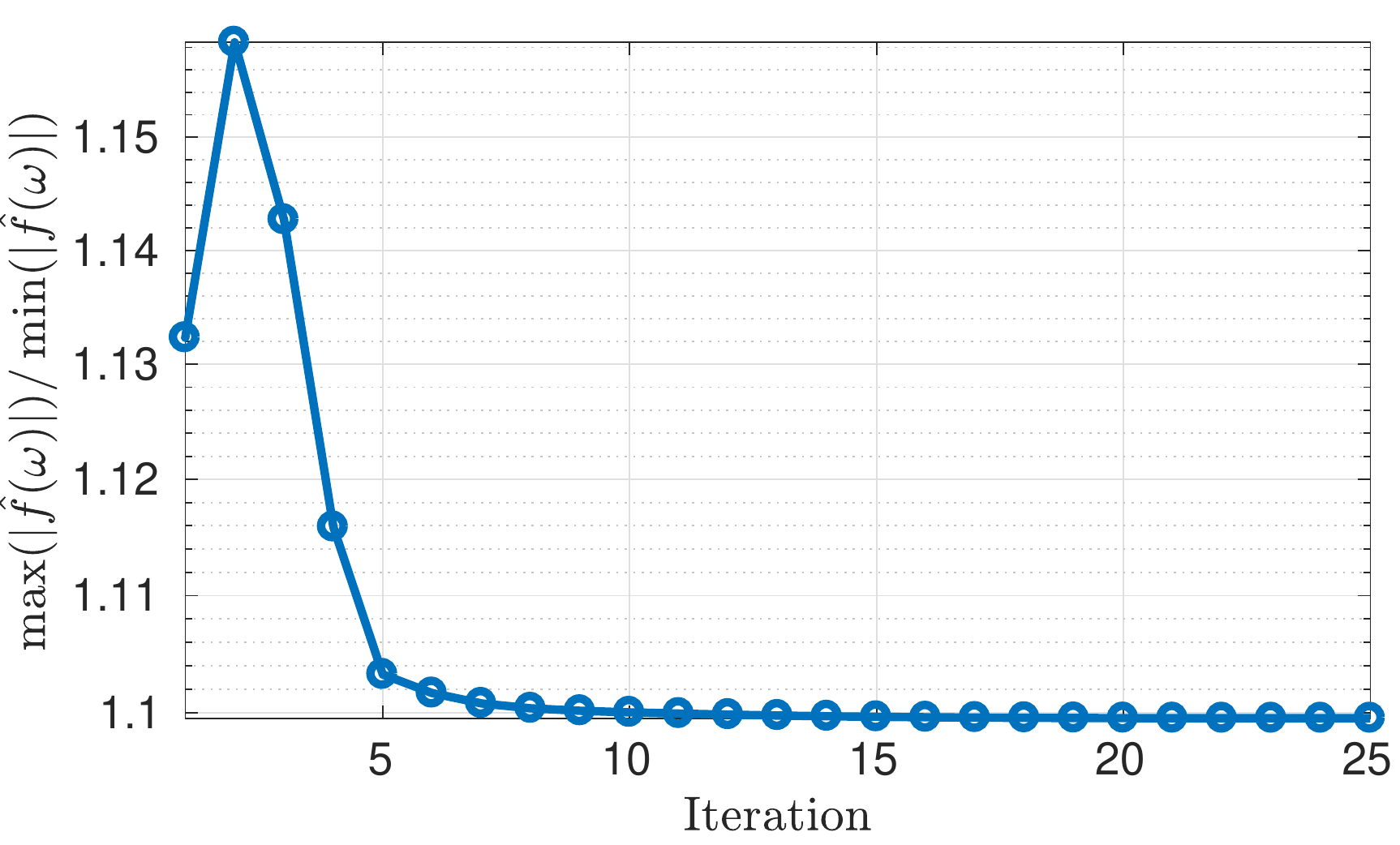} }
			
			\caption{Same as figure \ref{fig:amplitudesnotfiltered} with frequency normalization at each iteration.}
			\label{fig:amplitudesfiltered}
		\end{figure}

		A FIR filter of order 3000, with frequency resolution of $ \Delta \omega =\frac{1}{2\pi 15} \approx 0.2 $  is constructed. Although the filter order can be considerably smaller if the data is down-sampled, which becomes mandatory for large systems, due to the small size of this model this is an unnecessary complication.	Figure \ref{fig:amplitudesfiltered} shows that the filter regularizes the problem, yielding similar magnitudes for all frequency components.
		
\begin{figure}
			\centering
			\subfloat[Spectral amplitude]{\includegraphics[width=0.475\linewidth,trim={0 0 510 0},clip]{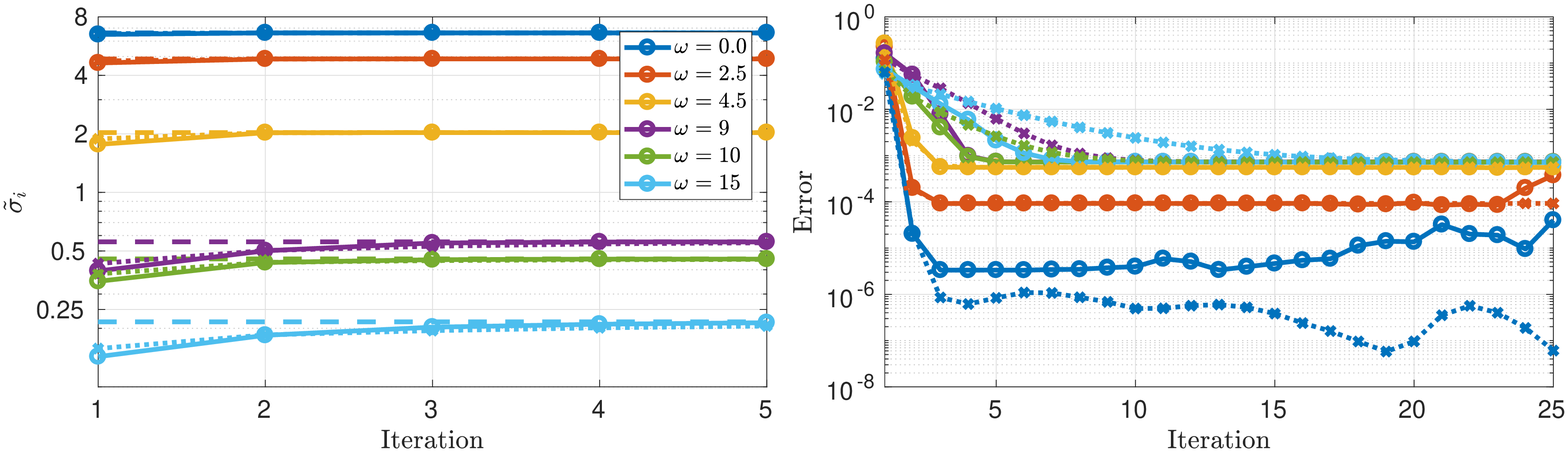} }
			\subfloat[Condition number]{\includegraphics[width=0.475\linewidth,trim={505 0 5 0},clip]{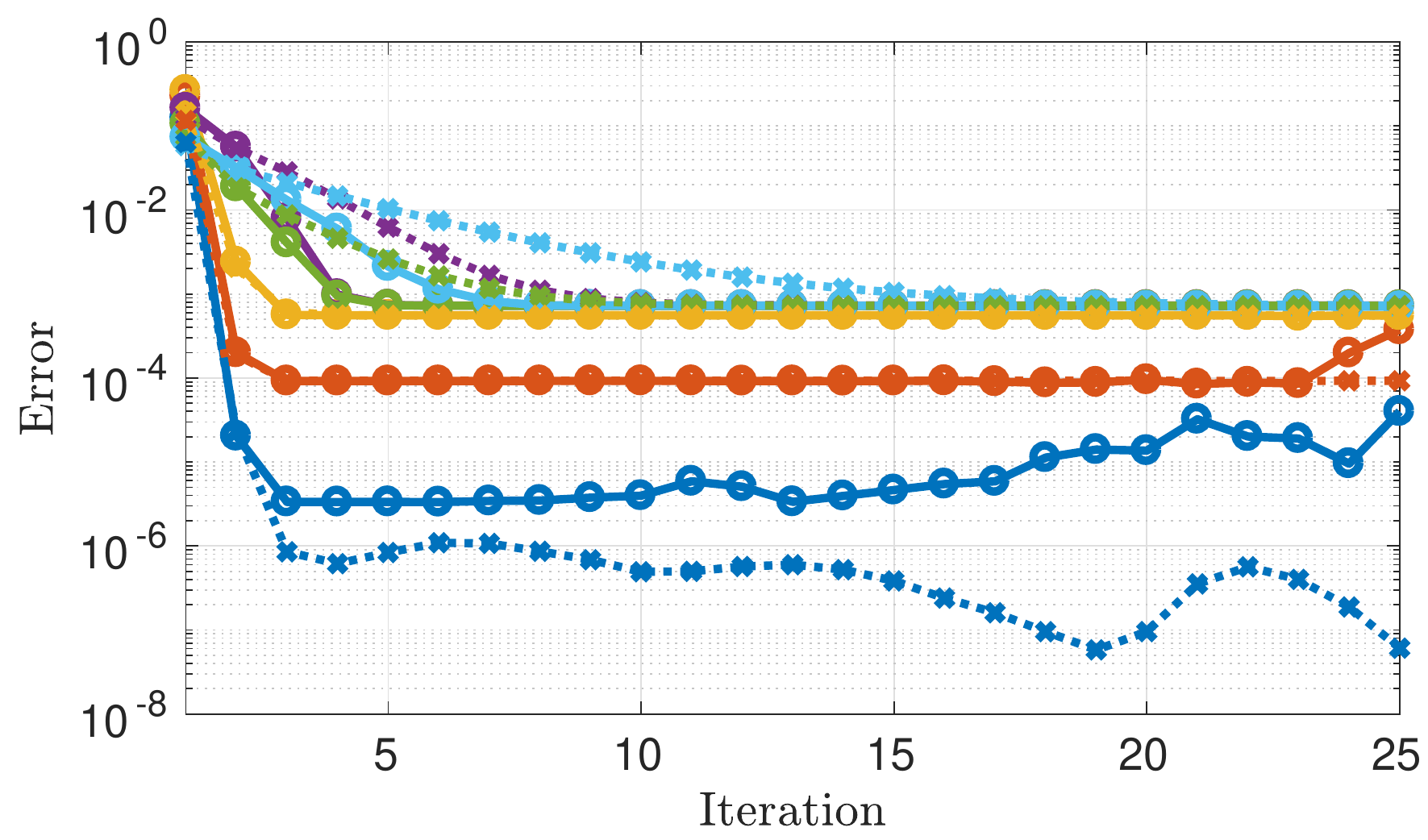} }
			
			\caption{Estimation of the leading resolvent gains, $\tilde \sigma$, (a) and errors (b) obtained with the power-iteration (dotted with crosses) and Arnoldi (solid with circles) algorithms. Error are defined as $ |\tilde \sigma_{1,i}-\sigma_1| $, with the application of a FIR filter after each iteration.}
			\label{fig:covergencegainsfiltered}
		\end{figure}

\begin{figure}
		
		\centering
		\includegraphics[width=\linewidth]{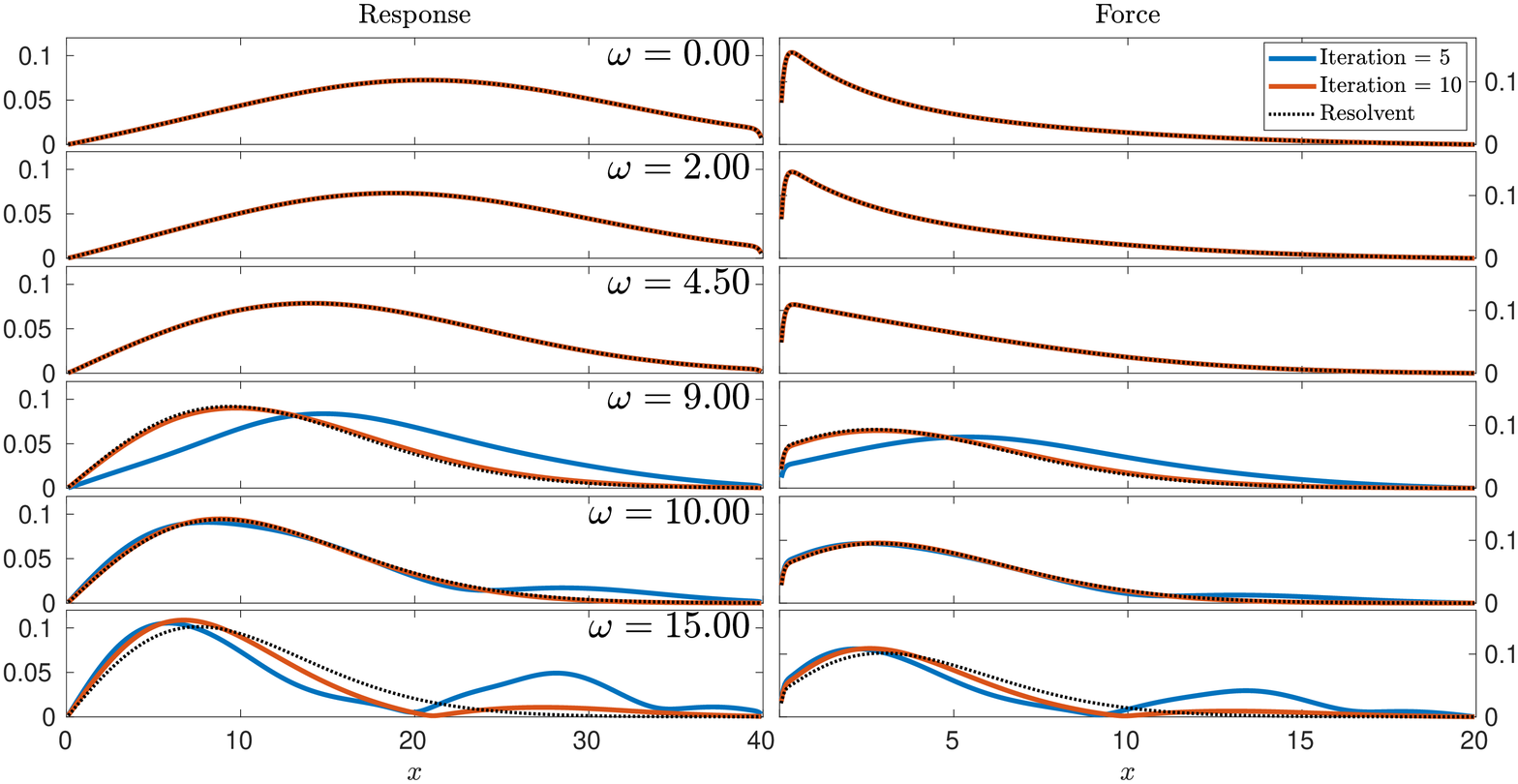}
		\caption{Absolute values of the estimated optimal force and response modes for different frequencies after 10 (blue) and 15 (red) iterations using the power-iteration algorithm. Black dashed lines correspond to the exact optimal modes.}
		\label{fig:optimalmodesfiltered}
	\end{figure}

\begin{figure}
 			\centering
 			\includegraphics[width=\linewidth]{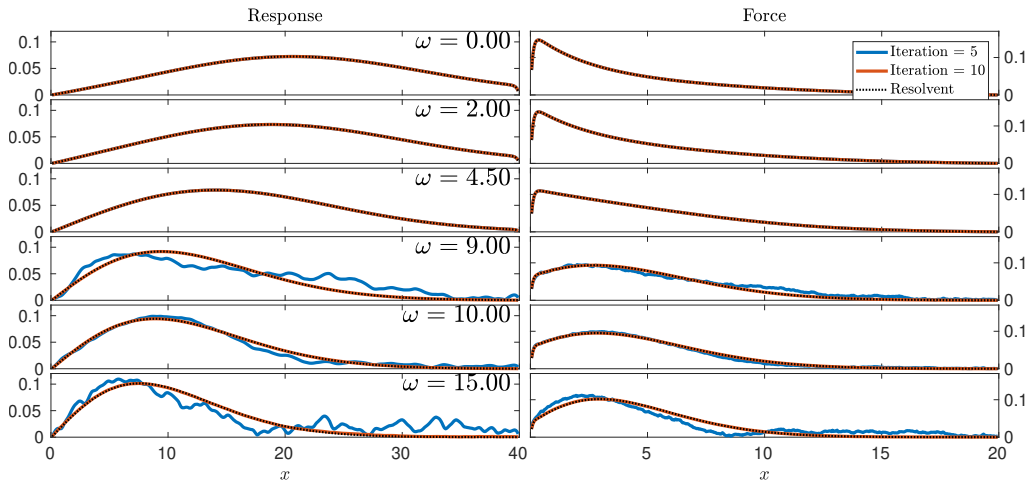}
 			\caption{Same as figure \ref{fig:optimalmodesfiltered} for the Arnoldi algorithm.}
 			\label{fig:optimalmodeskirlov}
 		\end{figure}

\begin{figure}
			\centering
			\includegraphics[width=\linewidth]{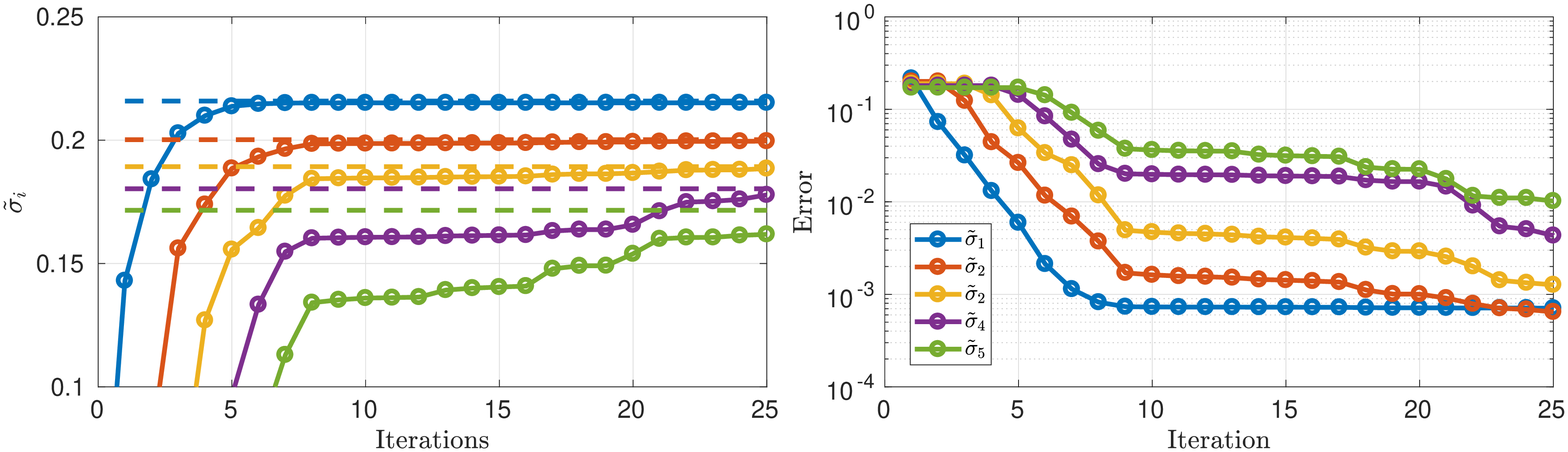}
			\caption{Convergence of the five leading gains for $ \omega=15 $. Error defines as in figure \ref{fig:covergencegainsfiltered}. }
			\label{fig:suboptimaw15}
		\end{figure}

		Figure \ref{fig:covergencegainsfiltered}  shows the convergence of gains observed with the power-iteration and Arnoldi algorithms. As discussed in \S~\ref{sec:discussion} the Arnoldi algorithm is not well suited for use with the TRM, however the small size of the problem studied here allows its application, providing a direct comparison between the two algorithms. The asymptotic error is related to the time marching scheme, and can decreased by reducing the time step.  The Arnoldi algorithm provides faster convergence when gain separation is small, as is the case for $ \omega=15 $.  The different convergence rates are associated with the different gain separation for each frequency, as expected from \eqref{eq:powerMethod}.  A comparison of modes computed with the power-iteration and Arnoldi algorithms, for the same number of iterations, is presented in figures \ref{fig:optimalmodesfiltered} and \ref{fig:optimalmodeskirlov}.  Figure \ref{fig:suboptimaw15}  shows the evolution of the gains computed for the first five modes at $\omega = 15$, illustrating the capability of the Arnoldi method to estimate sub-optimal modes, even when the gain separation is small.
		
\section{Resolvent modes of the incompressible flow around a parabolic body}\label{sec:parabolicBody}
	\subsection{Discretization and baseflow computation}
	The incompressible flow around a parabolic body is used to demonstrate both of the recommended approaches: the power iteration algorithm using TRM and the Arnoldi algorithm using SSRM.  
	
\begin{figure}
		\centering
		\subfloat[Full domain]{\includegraphics[width=.5\linewidth]{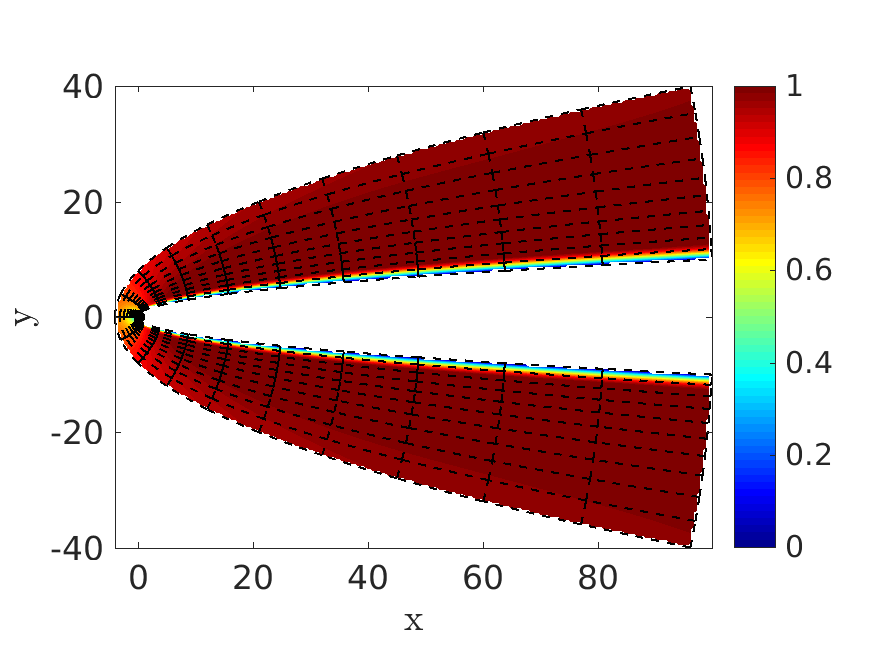}}
		\subfloat[Leading edge and boundary layer detail]{\includegraphics[width=.5\linewidth]{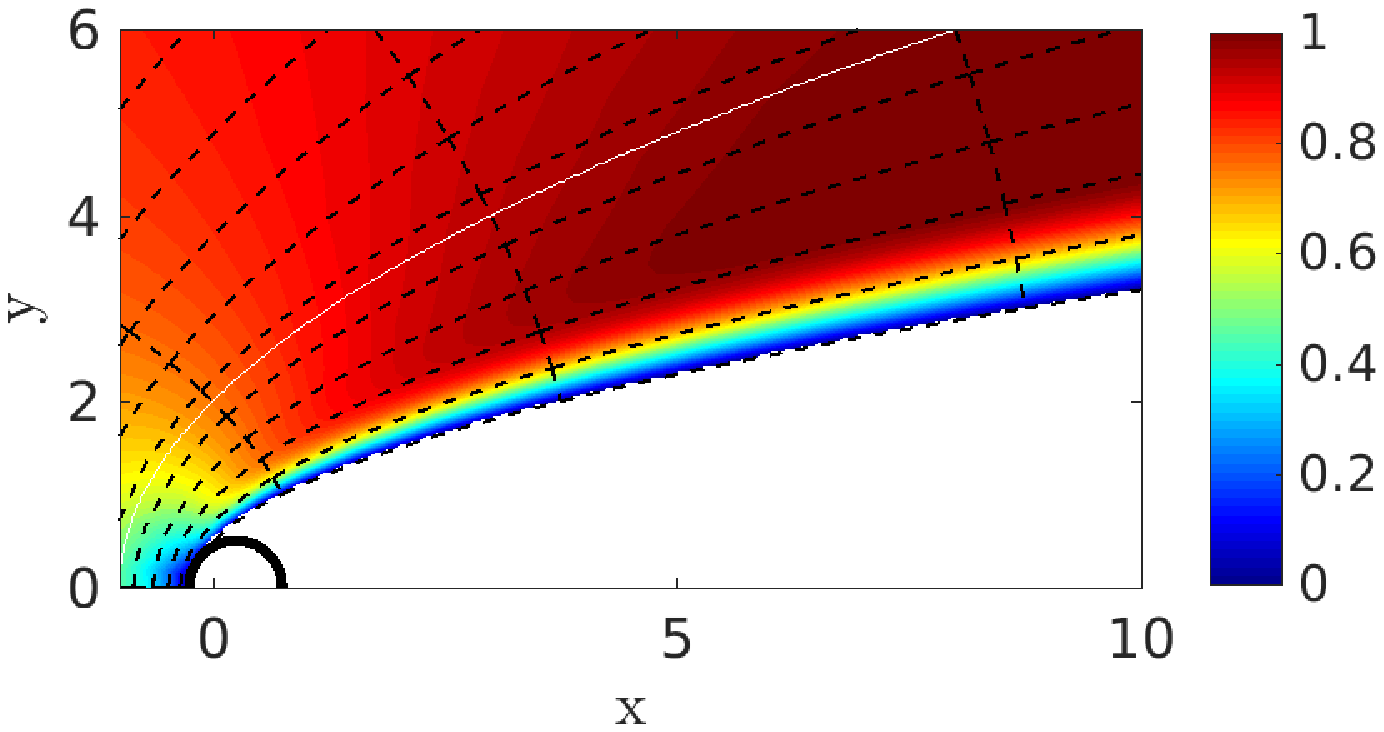}}
		\caption{Streamwise velocity field (colour scale) and element mesh for investigation of the flow around a parabolic body. The discretization uses $ 5^{th} $ order polynomials within each element. The while line delimits the region in which entries of $ \C $ are non-zero, and the black circle represents a circle with diameter of 0.5, tangent to the leading edge. }
		\label{fig:parabolic_baseflow}
	\end{figure}

	The Reynolds number based on the freestream velocity and leading edge curvature radius is 200.   The viscous base flow was taken as the stable laminar solution obtained by marching the Navier-stokes equations in time until the norm of the velocity time derivative becomes smaller than $ 10^{-8} $. No slip boundary conditions were applied at the body surface, with  outflow conditions on the right-most edge of the domain and inflow velocities obtained from an analytical solution of the potential flow, derived next, on the remaining boundaries. The mesh and the resulting baseflow are illustrated in figure \ref{fig:parabolic_baseflow}. The domain has a spanwise length of $10$ non-dimensional units, discretized with $6$ uniformly spaced spectral elements. Fifth-order polynomials were used for element discretization.
	
	Integration of the linear and non-linear Navier-Stokes equations were performed with the \emph{Nek5000} open-source code, which uses a spectral-element approach \citep{fischer1989parallel,fischer1998projection} based on $n^{th}$-order Lagrangian interpolants. The code contains routines to time march the direct and adjoint linearized Navier-Stokes equations, which were used to implement the methods proposed here. A validation of the resulting code is presented in appendix \ref{app:validation}. For the linearized problems, Dirichlet boundary condition were used on all boundaries. The model contains $ 1200 $ elements, corresponding to $ ~7.8\times10^{5} $  degrees of freedom.
	
	The geometry of the body suggests the use of parabolic coordinates for obtaining  the potential flow solution used as inflow conditions. The transformation between the Cartesian (x,y) and parabolic $ (\sigma,\tau) $ coordinates are given by
	\begin{align}
		x+\ii y = -(\sigma+\ii \tau)^2.
	\end{align}
	By inspection $ x = \tau^2-\sigma^2 $ and $ y=2\tau\sigma $. The solid surface is located at $ x= y^2 + 1/4 $, corresponding to a constant value of $ \sigma $,  $\sigma_0=0.5 $.
	The flow streamfunction is obtained by a solution of the Laplace equation
	\begin{align}
		\nabla^2_{\sigma,\tau} \psi = 0,
	\end{align}
	with the following boundary conditions: no penetration condition at the body surface, $ \psi=0 $ at $ \sigma=\sigma_0 $; convergence to the uniform, right moving flow, away from the body, $ \psi=2\sigma\tau=y $ at $\sigma\to\infty $. The potential is then written as $ \psi=2(\sigma-\sigma_0)\tau $, which clearly satisfies the boundary conditions.

	\subsection{Computation of resolvet modes}
	
 	To focus the response of the flow within the boundary layer, a diagonal matrix $ \C $ with $ 1 $ in the region close the body, indicated by the white line in figure \ref{fig:parabolic_baseflow}, and zero otherwise was used, minimizing the excitation of free-stream vortices, as described by \cite{nogueira2020resolvent}.  No restrictions on the force terms was imposed, i.e. $ \B=\I $.  As three-dimensional simulations of the linearised system are performed, resolvent modes for all spanwise wavenumbers matching the domain size are obtained simultaneously.

	 For the TRM, integration was carried out until the norm became $ 10^{-3} $ of the maximum obtained during the run. Flattening filters were obtained based on  96 frequencies non-uniformly spaced between 0 and $ 0.5 \pi $, and constructed with order $87$   to obtain frequency resolution of $ 0.2\pi $ , and designed with a cut-off frequency of $ \pi  $. For the SSRM 140 time units were used for vanishing of the initial conditions, an interval for which a random initial perturbation reached a norm of $ 10^{-3} $, with results obtained for frequencies $ \omega_j=0.04\pi j $, with  integer $ j $ between 0 and 25.
 	
\begin{figure}
		\centering
		\subfloat[Frequency dependency of gains]{\includegraphics[width=\linewidth]{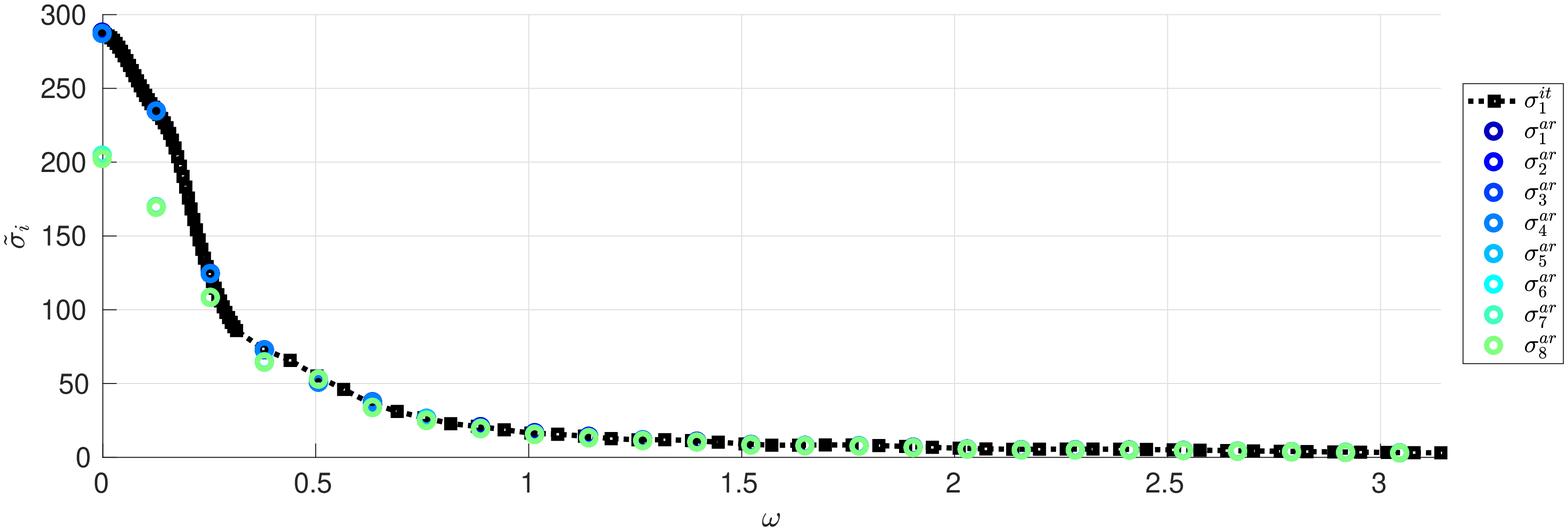}}
		
		\subfloat[Gain convergence for $ \omega=0.00$] {\includegraphics[width=\linewidth,trim={0 0 0 
				19px },clip] {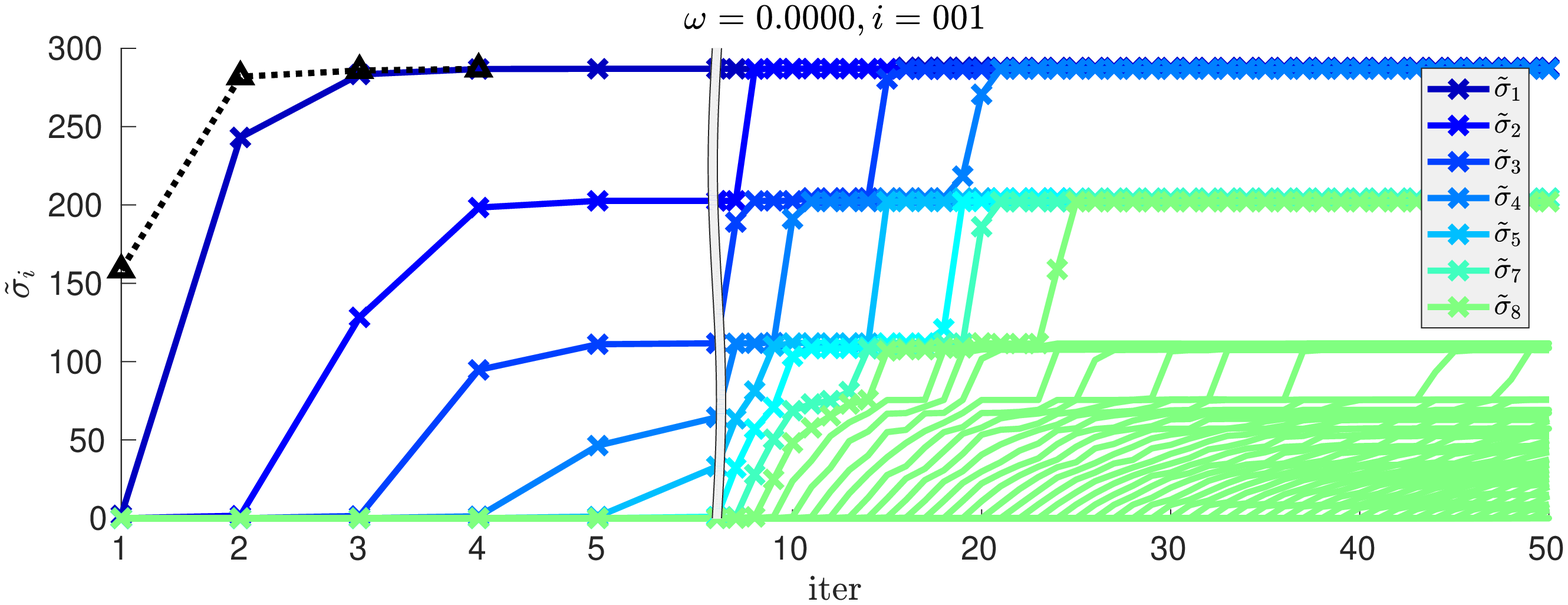}}

		\subfloat[Gain convergence for $ \omega=0.25$]{\includegraphics[width=\linewidth,trim={0 0 0 
				19px },clip]{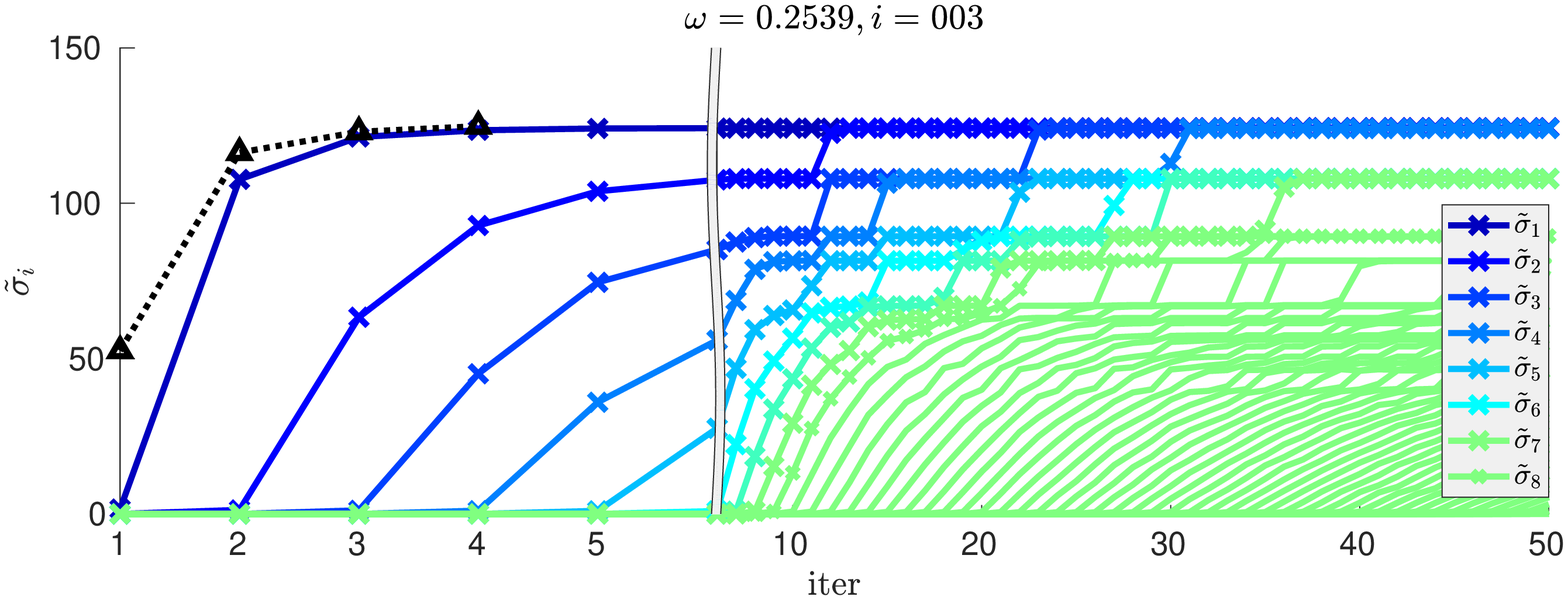}}
		\caption{Leading gains for the parabolic body: (a) gains as a function of frequency, (b) and (c) gains convergence with iteration count. Results from the SSRM using the Arnoldi algorithm in coloured lines, and results from the TRM with the power iteration algorithm in black. }
		\label{fig:parabolic_gains}
	\end{figure}

\begin{figure}
		\centering
	
		\subfloat[$\omega=0.00$]{\includegraphics[width=\linewidth,trim={0 0px 0 0},clip   ]{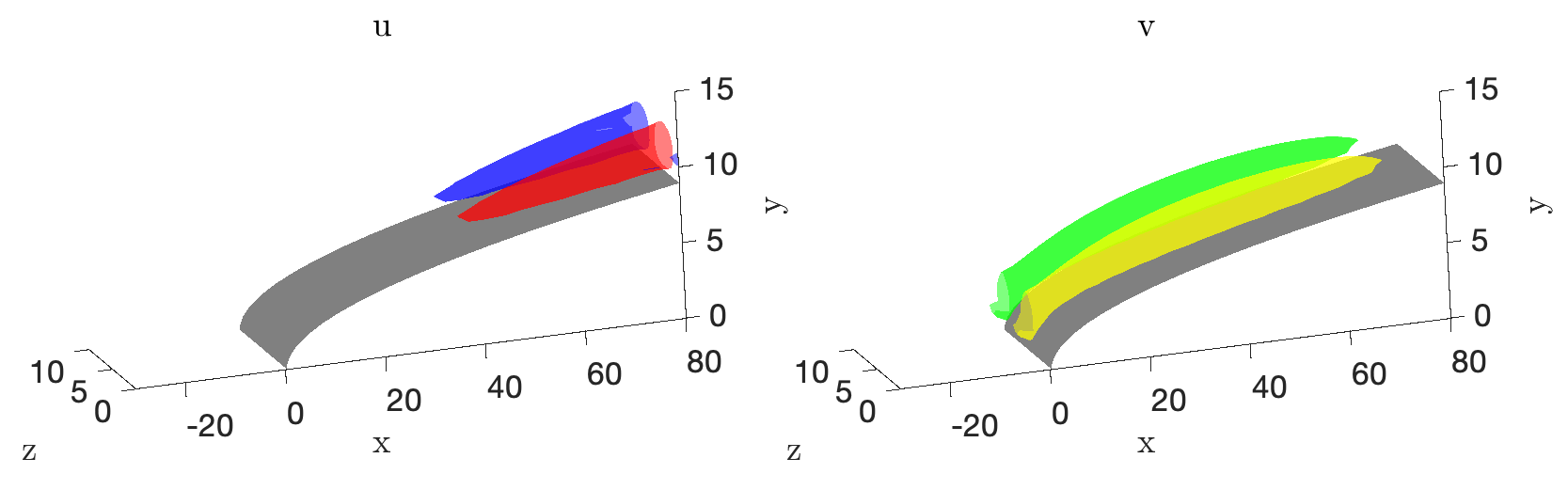}}
		
		\subfloat[$\omega=0.13$]{\includegraphics[width=\linewidth,trim={0 0px 0px 0},clip   ]{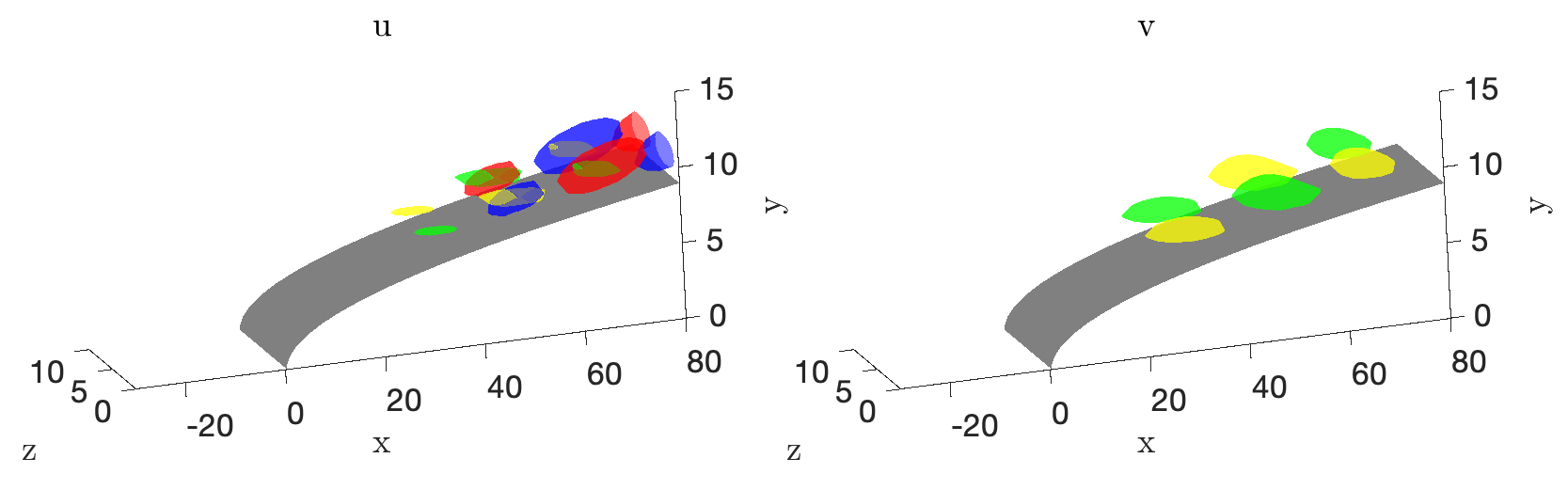}}

		\subfloat[$\omega=0.25$]{\includegraphics[width=\linewidth,trim={0 0px 0px 0},clip   ]{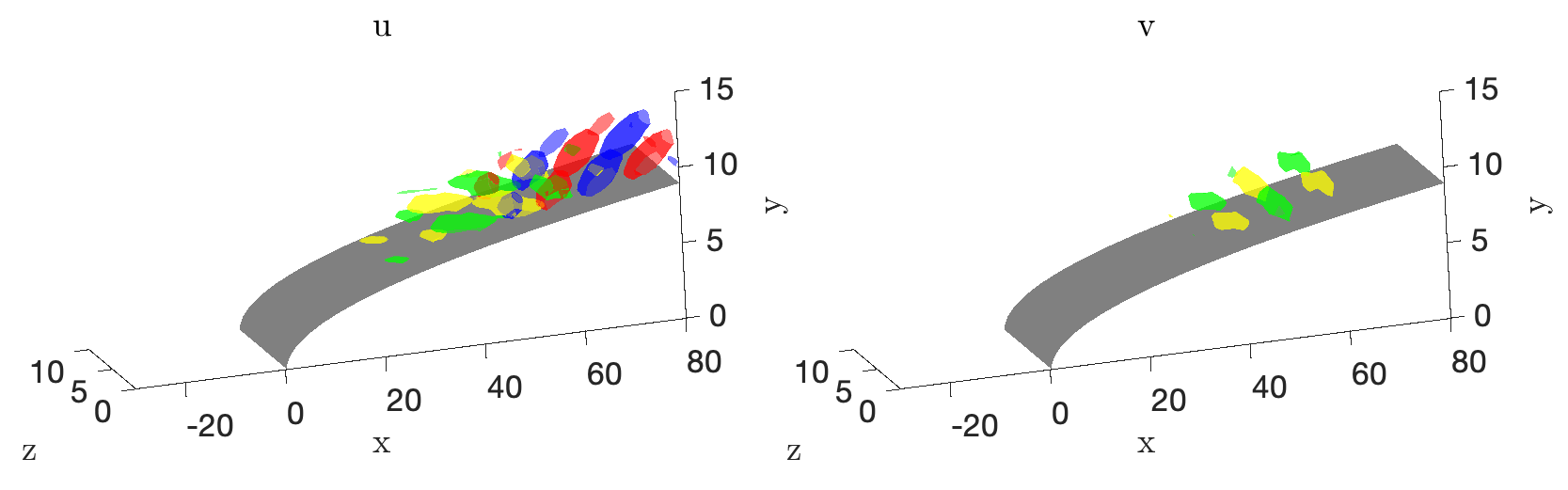}}
		
		
		\caption{Real part of optimal force (green and yellow) and response modes (red and blue) for the flow around a parabolic body. On each subplot forces and responses in the $ x $ (left) and $ y $ (right)  directions are shown.}
		\label{fig:parabolic_optModes}
	\end{figure}

\begin{figure}
		\centering
		
		\subfloat[Optimal modes]{\includegraphics[width=.5\linewidth,trim={0 0px 0 0},clip   ]{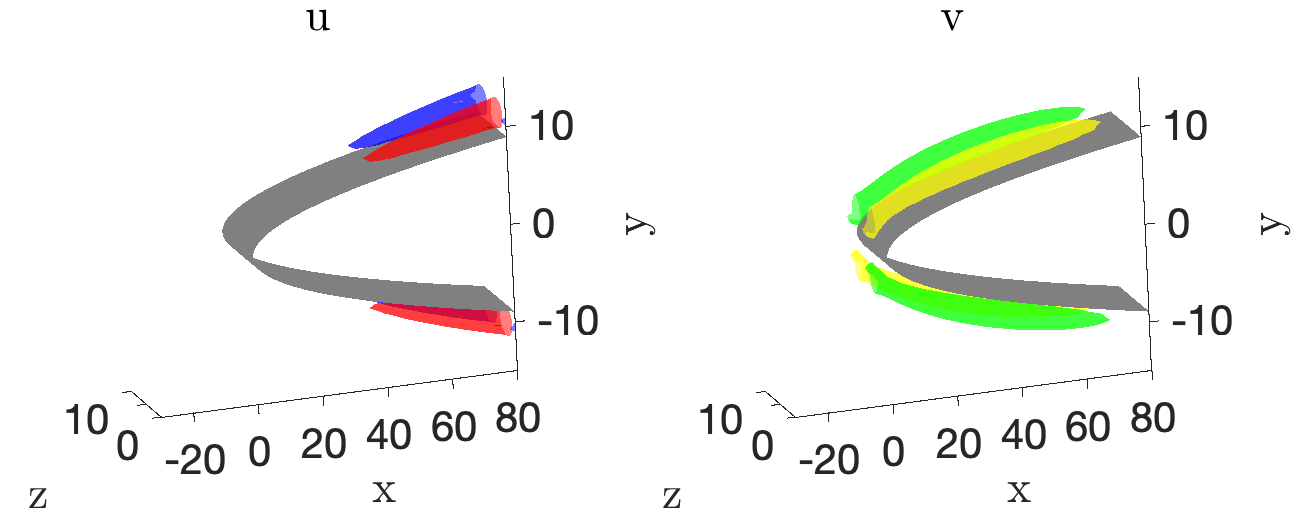}}
		\subfloat[First suboptimal modes]{\includegraphics[width=.5\linewidth,trim={0 0px 0 0},clip   ]{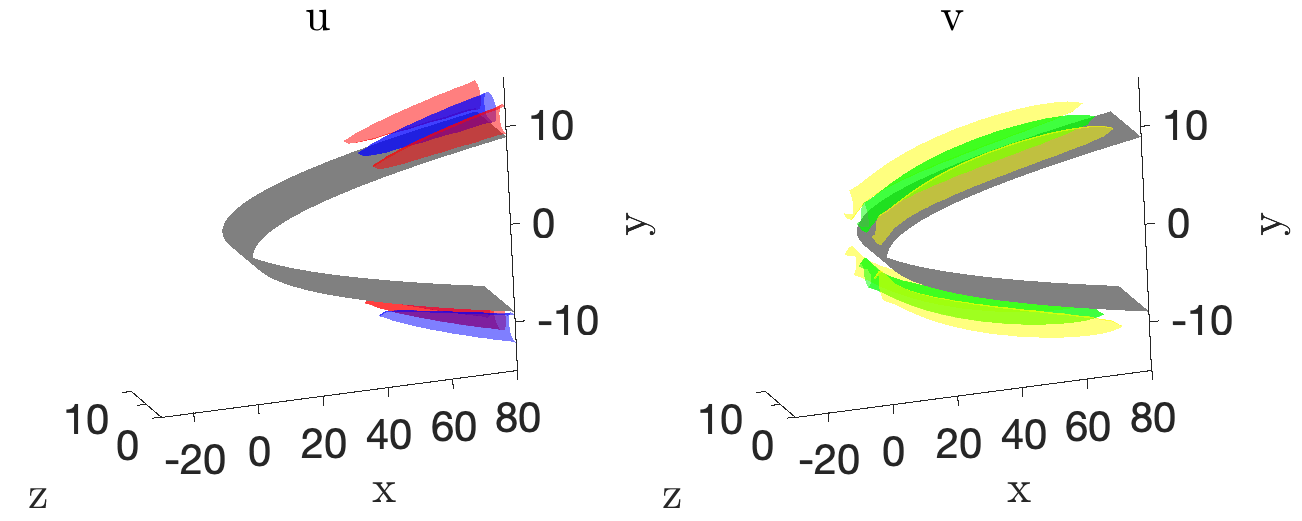}}
		
		\subfloat[Second suboptimal modes]{\includegraphics[width=.5\linewidth,trim={0 0px 0 0},clip   ]{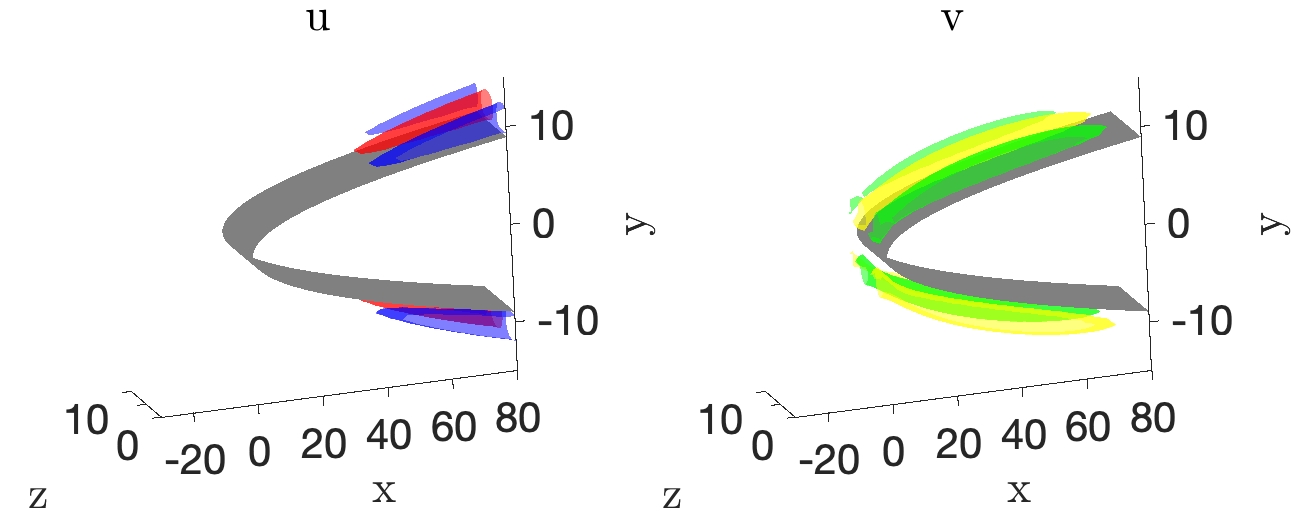}}
		\subfloat[Thrid suboptimal modes]{\includegraphics[width=.5\linewidth,trim={0 0px 0 0},clip   ]{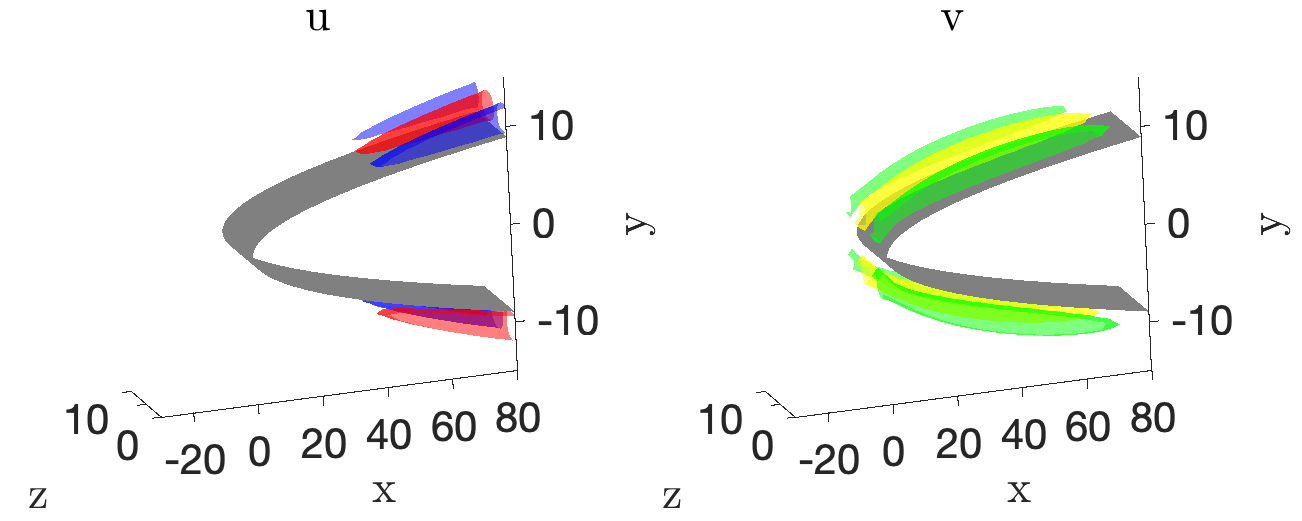}}
		
		\subfloat[Fourth suboptimal modes]{\includegraphics[width=.5\linewidth,trim={0 0px 0 0},clip   ]{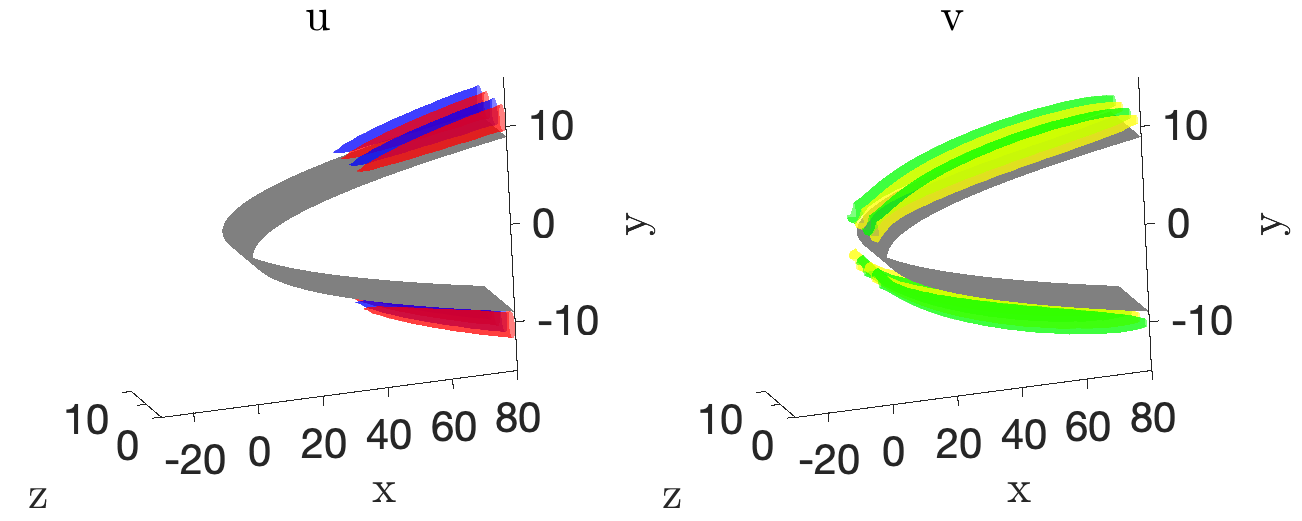}}
		\subfloat[Fifth suboptimal modes]{\includegraphics[width=.5\linewidth,trim={0 0px 0 0},clip   ]{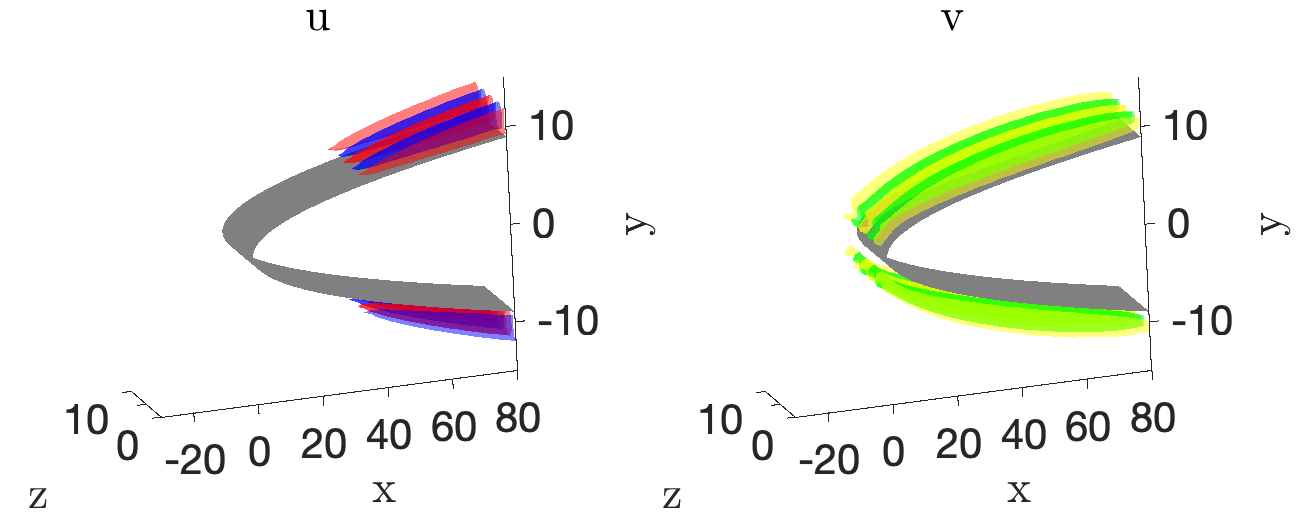}}
		\caption{Same as figure \ref{fig:parabolic_optModes} for optimal and  suboptimal modes at $ \omega=0.00 $.}
				\label{fig:parabolic_suboptModes}
%
%
%
	\end{figure}

 	Figure \ref{fig:parabolic_gains} shows gains as a function of frequency, and their convergence with iteration count. In total 4 iteration were performed with the transient-state method, and 50 with the SSRM. The highest gains are found at $\omega = 0$, with responses dominated by stream-wise velocity components, with force terms exciting streamwise vortices. The mechanism is consistent with the lift-up effect in transitional boundary layers \citep{monokrousos2010global}. At higher frequencies this mechanisms becomes less efficient, and free-stream structures near the wall dominate the system. It can also be noted that there are four modes  which converge to approximately the same value. These consists of $ cosine $ and $ sine$ components in the $ z $ direction, which should provide exactly the same gains, as $ z $ is a homogeneous direction, and to symmetric and asymmetric modes with respect to the $ x-z $ plane, which should have similar gains if there is little interaction between both sides of the body. 	Numerical errors from spatial-temporal discretization, remaining transient effects (SSRM), or truncation of the time series (TRM), can generate small differences between $ cosine $ and $ sine $ modes, and mask the distinction of symmetric and asymmetric modes. Here the distinction between these gains is negligible, and they effectively span an optimal subspace. Figure \ref{fig:parabolic_optModes} shows the upper half-domain of the leading modes for different frequencies. Figure  \ref{fig:parabolic_suboptModes}  shows suboptimal modes for $ \omega=0 $. Vectors describing the optimal subspace where chosen as to better represent the symmetries of the problem.
	
\begin{figure}
		\centering
		\includegraphics[width=\linewidth]{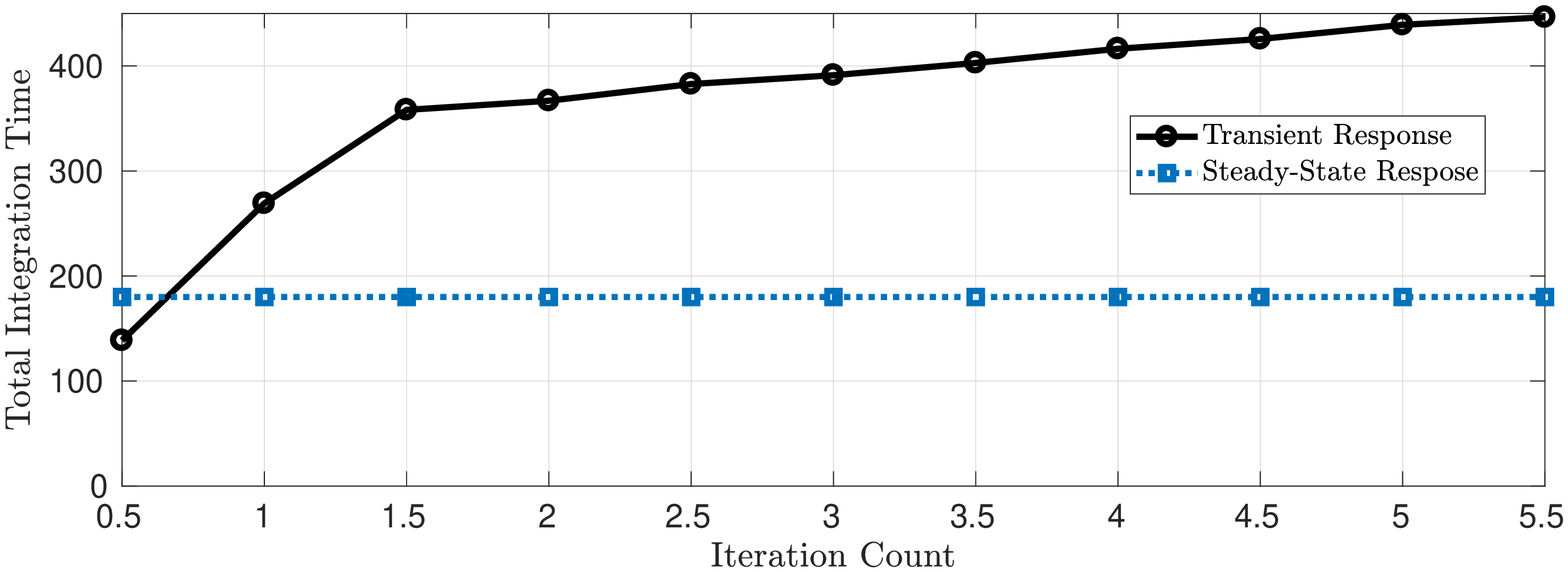}
		\caption{Total integration time using the different approaches. Half-integer/integer values reefer to the direct/adjoint runs. }
		\label{fig:parabinttime}
	\end{figure}

	Figure \ref{fig:parabinttime} shows a comparison with the integration time required by each method. On the SSRM, an integration of 140 time units was used for eliminating transient effects, with another $40$ units required to compute resolvent modes for the desired set of frequencies. All integrations have thus the same time length. Note however that the time needed to characterize the frequency responses increases with the inverse of frequency resolution. The TRM shows increasing integration times for each iteration, reaching $ \approx 3.5 $ times the initial time integration length at the last iteration. this cost, however, does not scale with the frequency discretization.

	The optimal gains reported here are considerably larger than previously receptivity mechanism reported in the literature \citep{haddad1998boundary}, and for other similar configurations \citep{shahriari2016acoustic}. Although such difference is not surprising, given that these studies focus on the receptivity of Tolmien-Schlichting (TS) waves to incident acoustic waves, and here we are focusing on receptivity to distributed forces within the domain.
	It is beyond the scope of this work to perform a full investigation on the relevance of the receptivity of vortical disturbances to the dynamics of the flow, but the results presented here indicate that they might be a important mechanisms for transition.
	Boundary layers subject to significant levels of free-stream turbulence have a transition previously demonstrated to be dominated by streaks, \citep{matsubara2001disturbance}, in a process that is affected by the leading edge geometry \citep{nagarajan_lele_ferziger_2007}, which further support that receptivity to distributed forces can be an important ingredient to transition.
	
\section{Conclusions} \label{sec:conclusions}	 

	Two novel methods to obtain resolvent modes and gains were presented and allow the computation of gains and modes for several frequencies simultaneously.
    The transient-response method (TRM) allows for a fine frequency discretization for the computation of leading modes and gains.  The steady-state response method (SSRM)  allows suboptimal and high frequencies to be computed with the use of the Arnoldi algorithm. To the best of the authors knowledge, this is the first implementation of the Arnoldi algorithm for the computation of resolvent modes and gains using a matrix-free approach. 
	
	Convergence trends where shown for a linearized Ginzburg-landau system, where gains and modes where computed with the proposed method, and with a direct singular-value decomposition. Geometric convergence rates were observed. An implementation within an incompressible solver was used to compute gains and modes on the flow around a parabolic body, for  a setup consisting of  $\approx 0.8 $ million degrees of freedom. Optimal and suboptimal gains and modes were reported.  
	
	For large system,  the dominant cost is the time marching of the direct- and adjoint-linearized equations. Results for multiple frequencies are obtained, leading to a cost reduction of at least an order of magnitude when compared to previous time marching methods for computing resolvent modes \citep{monokrousos2010global}.  
	

	\appendix
	
	\section{Algorithms} 
	\label{app:algorithms}
	Here an overview of the iterative algorithms used in this work to compute singular values is presented. We focus on practical aspects which will be necessary for the development of the codes and methods used. For a more complete review of algorithms applicable to large system we referrer the readers to \cite{saad2003iterative}.

	The leading singular value and associated modes can be obtained via the power-iteration algorithm, for which convergence can be easily derived analytically. First, define a test vector
	\begin{equation}\label{key}
	\hat\f_0 = \sum_{i=1}^{\n_f} a_i \hat{\vv{\mathcal{V}}}_i.
	\end{equation}
	Using \eqref{eq:systemFreqDomain} and \eqref{eq:adsystemFreqDomain} to form an iterative scheme
	\begin{align} \label{eq:iteration}
	\hat \w_n=&  \Ry^\dagger \Ry \hat \f_n  	
	\end{align}
	and choosing $ \hat \f_n = \hat \w_{n-1} $, for non-zero $ a_1 $, the term $ \hat \f_n $ can be written as 
	\begin{equation}\label{eq:powerMethod}
	\hat \f _{n}= (\Ry^\dagger \Ry)^n \hat \f_0 = \V \Sigma^{2n} \V^\dagger \hat \f_0 = a_1 \sigma_1^{2n} \left( \vv{\mathcal{V}_1} + 
	\sum_{i=2}^{\n_f} \frac{a_i}{a_1}\frac{\sigma_i^{2n}}{\sigma_1^{2n}} \hat {\vv{\mathcal V}}_i\right).
	\end{equation}
	Assuming $ \sigma_1 > \sigma_2 $, for large $ n $, $ \hat \f_n/\sigma_1^{2n}  \approx  \vv{\mathcal V}_{1} $, i.e. it converges to the leading force mode, and the leading gain can be estimated as 
	\begin{align}
	\tilde \sigma_{1,n}(\omega)  &= \sqrt{\lvert\lvert \hat \f_ n(\omega)\rvert\rvert \over  \lvert\lvert \hat \f_{n-1}(\omega)\rvert\rvert } \label{eq:tildeSig}.
	\end{align}
	
	Asymptotically, the power iteration algorithm has a geometrical convergence rate, with error reducing by a factor of  $ (\sigma_1/\sigma_{2})^{2} $ on each iteration .  In general, $ \hat \f_n $ converges to the subspace spanned the first $ m $ force modes with a rate given by $ \sigma_m/\sigma_{m+1} $. This is particularly relevant if $ \sigma_1=\dots=\sigma_m $. In such case $ \hat \f_n $ converges to one singular mode in the $m$-dimensional subspace, with rate $ \sigma_1/\sigma_{m+1} $. This ratio will be refereed to as \emph{gain separation}.
	
	If gain separation is small, i.e. close to 1, many iterations might be needed for convergence of the power-iteration algorithm. Alternatively, gains can be estimated based on a low-rank representation of $ \Ry^\dagger \Ry $ on subspace spanned by a sequence of vectors $ \hat \f_n $, with $ 1\le n \le m $. Given a sequence of $ \hat \f_n $ and $ \hat \w_n $ satisfying 
	\begin{equation}\label{eq:RRfpi}
	(\Ry^\dagger \Ry )\hat \f_n  = \hat \w_n.
	\end{equation}
	Using $ \hat \f_n=\hat \w_{n-1} $, the subspace spanned by $ \hat \f_n $ is  the \emph{Krylov subspace}. From \eqref{eq:powerMethod} it is clear that for large $ n $ this subspace asymptotically includes the leading force mode. 
	Defining $ \F = \left[ \hat \f_1, \dots, \hat \f_n \right] $ and $ \W = \left[ \hat \w_1, \dots ,\hat \w_n \right] $, \eqref{eq:RRfpi} can be written as
	\begin{align} \label{eq:RRFi}
	(\Ry\Ry^\dagger) \F = \W,
	\end{align}
	From a QR decomposition, $ \F=\mm Q_{F}\R_{F} $, where $ \mm Q_{F} $ is a unitary matrix and $ \R_{F} $ is upper triangular, \eqref{eq:RRFi} can be re-written as 
	\begin{align}\label{eq:arDec}
	(\Ry\Ry^\dagger) \mm Q_{F} = \W \R_{F}^{-1} = 
	\mm Q_{F} \underbrace{\mm Q_{F}^\dagger \W \R_{F}^{-1}}_{\mm H} + \left(\I -\mm Q_{F}\mm Q_{F}^\dagger \right)\W \R_{F}^{-1}.
	\end{align}
	Since $ \mm Q_F $ forms a orthonormal basis for the space spanned by $ \hat \f_n $, a low-rank representation of $ \Ry^\dagger\Ry $ in  this space is given by
	\begin{align}
	\mm Q_{F}^\dagger(\Ry\Ry^\dagger) \mm Q_{F} = 
	\mm H, 
	\end{align}
	where $ \mm Q_{F}^\dagger \mm Q_{F} = \I $ was used. The components of $ \hat \f_n $ orthogonal to this space are restricted to the right-most term. The eigenvalues and vectors of $ \Ry^\dagger \Ry $ can then be estimated from an eigen-decomposition of $ \mm H $
	\begin{align}
	\mm H \Psi = {\mm \Gamma}^2 \mm\Psi,
	\end{align} 
	where $ \Gamma $ is diagonal with entries $ \gamma_1\ge\gamma_2\ge\dots\ge \gamma_m $, and $ \Psi $ is unitary, with the j-th row represented by $ \psi_j $. Force and response modes, and gains associated with them, can be estimated as,
	\begin{align}\label{key}
	\tilde {\vv {\mathcal{V}}}_i& = 	   \F \psi_i, &
	\tilde {\vv {\mathcal{U}}}_i& = 	\gamma_i^{-1}\Ry\F \psi_i, &
	\tilde \sigma_1 & =  \gamma_1,
	\end{align}
	where $ \tilde \cdot $ represent the estimated values.
	
	If $ \hat \f_n=\hat \w_{n-1} $, the right-most term can be shown to have rank one, \eqref{eq:arDec} is an \emph{Arnoldi factorization} of $ (\Ry\Ry^\dagger) $, and the matrix $ \mm H $ is Hessenberg and Hermitian. For large $ n $, $ \hat \f_n $ approximates the leading force mode, and thus the last vectors in the sequence become approximately linearly dependent. This leads to an ill-conditioning of the inversion of $ \R_{F} $. To avoid this problem, the Arnoldi algorithm can be used.  It consists of using as  $ \hat \f_n $ the component of $ \hat\w_{n-1} $ which is orthogonal to all previous forces $ \hat \f_j $, with $ j<i-1 $. This can be obtained as 
	\begin{equation}\label{eq:freqOrt}
	\hat \f_n = \hat \w_{n-1} - \mm F_{0,..,n-1} \hat {\vv \theta}  ,
	\end{equation}
	where
	\begin{align}
	\mm F_{0,..,n-1}&=\left[\hat \f_0,\dots,\hat \f_{n-1} \right], \\
	\hat {\vv \theta}_i &=  \left( \mm F^H_{0,..,n-1} \mm W_f \mm F^+_{0,..,n-1}  \right)^{-1} \mm F^+_{0,..,n-1} \W_f \hat \w_{n-1}.
	\end{align}
	
\begin{figure}
		\centering
		\includegraphics[width=\linewidth]{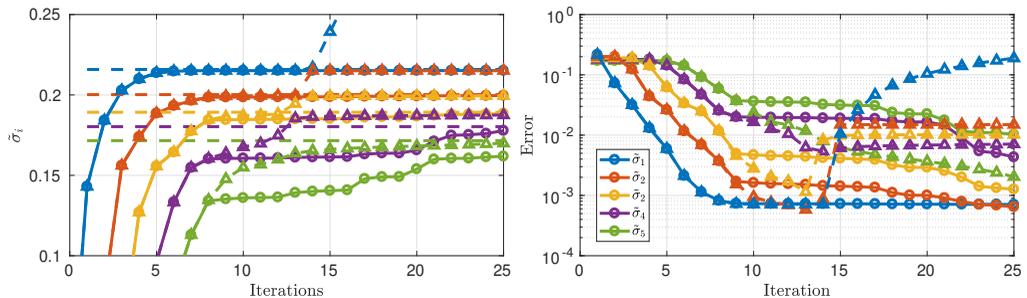}
		\caption{Same as figure \ref{fig:suboptimaw15}, with the addition of the Krylov-subspace algorithm (triangles). }
		\label{fig:suboptimaw15_krylov}
	\end{figure}

	Note however that such ill-posedness only occurs once $ \hat \f_n $ has converged to the leading force mode, and thus if only the leading gain and modes are of interest either the inversion of $ \R_{F} $ is well conditioned, or the leading modes and gains can be accurately obtained from the power-iteration algorithm. The use of the Arnoldi algorithm is therefore only necessary if sub-optimal gains and modes are of interest.
	 Figure \ref{fig:suboptimaw15_krylov} reproduces figures \ref{fig:suboptimaw15}  adding the trend observed with the Krylov algorithm. Krylov-subspace has the same convergence trend of the Arnoldi algorithm for the first iterations, and accurately captures the optimal gain before the algorithm becomes ill conditioned, and thus is a viable alternative to accelerate computation of the leading mode if convergence rates are low due to small gain separations. 
	 
	\section{Temporal filter overview}\label{app:filters}
		Finite impulse response (FIR) filters are a class of filters which can be applied to uniformly spaced time sequences. Given a signal sampled  $ x(j \Delta t ) $ and a filter $ \phi(j \Delta t) $, the filtered signal $ x'(j \Delta t) $ is obtained as
		\begin{equation}\label{key}
			x' = \phi * x,
		\end{equation}
		where $ * $ is a discrete convolution. In the frequency domain the filtered signal can be expressed as
		\begin{equation}\label{key}
		\hat x' = \hat \phi  \hat x.
		\end{equation}
		If $ \phi $ is bounded, that is $ \phi(t)=0 $ for  sufficiently large $ |t| $, the filter is a \emph{finite impulse response} filter. The filter order is related to the number of points at which $ \phi(j \Delta t) $ is non-zero: a $ n $-th order filter has $ n+1 $ non-zero points. Higher-order filters have better frequency resolution, with low-order filters having resolution only for frequencies close to the Nyquist frequency, $ \omega_{nyq} = \pi/(\Delta t) $. 
		
\begin{figure}
			\centering
			\subfloat[Time-domain signals]{\includegraphics[width=0.5\linewidth]{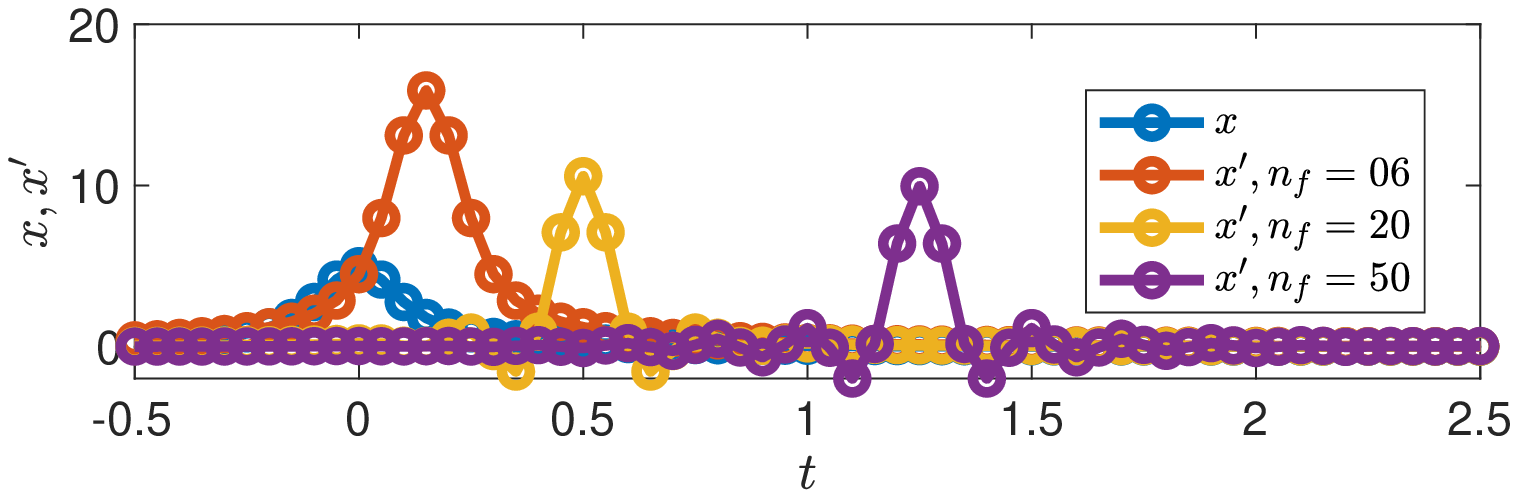}}
			\subfloat[Frequency-domain signals]{\includegraphics[width=0.5\linewidth]{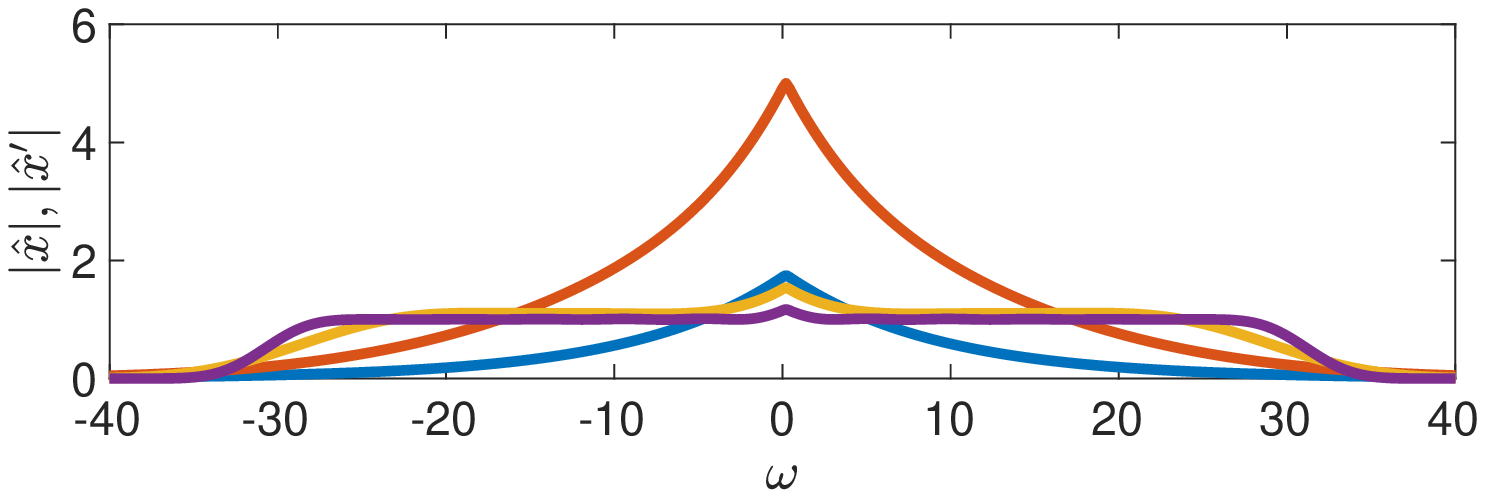}}
			
			\subfloat[Filter ]{\includegraphics[width=0.5\linewidth]{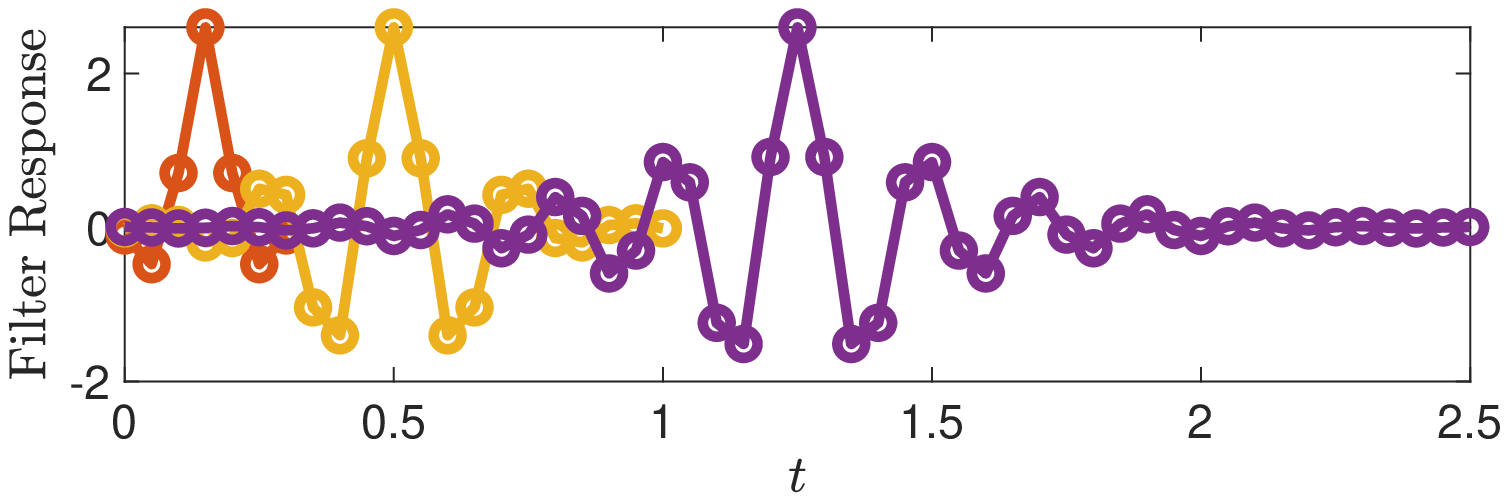}}
			\subfloat[Filter Spectral Response]{\includegraphics[width=0.5\linewidth]{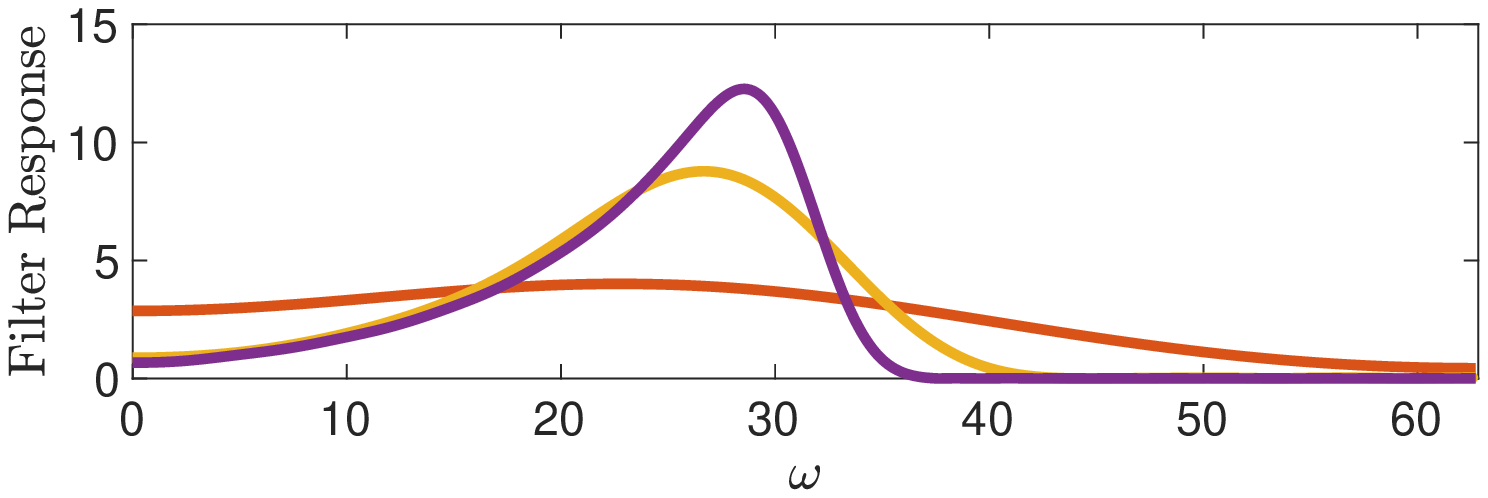}}
			\caption{Effect of filter order ($ n_f $) on flattening a signal spectra.}
			\label{fig:filter1}
		\end{figure}

		Figure \ref{fig:filter1} shows an example of flattening a signal content with FIR filters obtained with Matlab \emph{fir2} filters, which produces a filter that best approximates the desired amplitude gains.
		The filter was designed as to flatten the spectra for frequencies lower than $ 10\pi $ and to reduce amplitudes for higher frequencies. This is used in the steady-state-response method to reduce aliasing. Note that there is signal delay, which increases with the filter order. The \emph{fir2} function creates filters that have a linear phase shift corresponding to a time delay of $ n \Delta t /2 $, which can be easily compensated for.
		
\begin{figure}
			\centering
			\subfloat[Time-domain signals]{\includegraphics[width=0.5\linewidth]{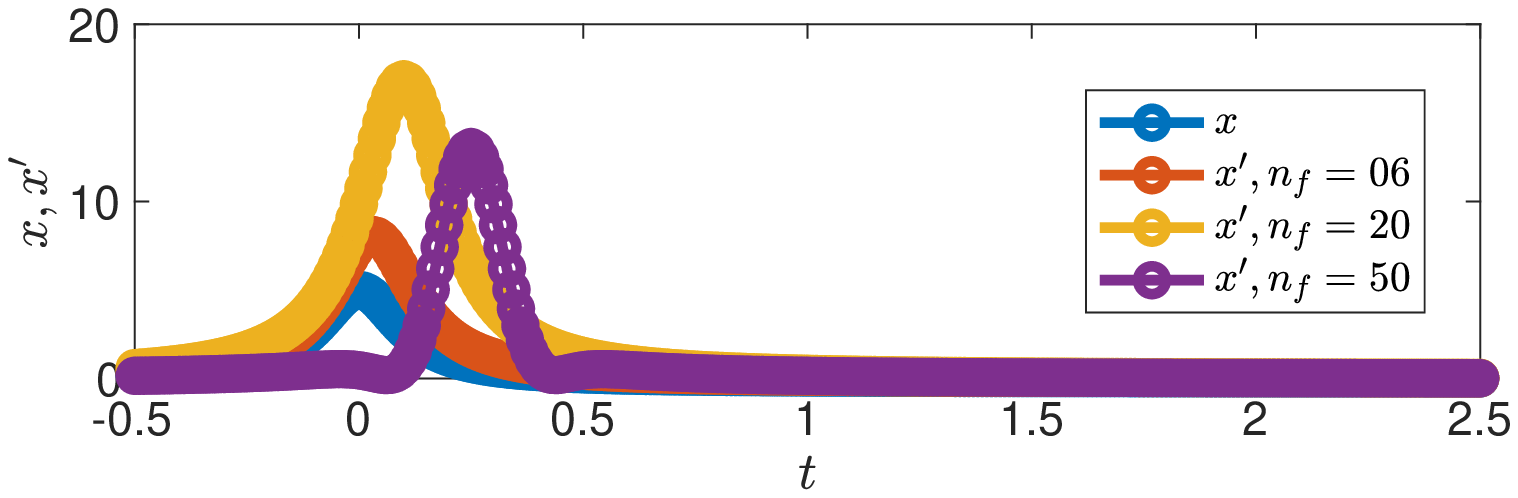}}
			\subfloat[Frequency-domain signals]{\includegraphics[width=0.5\linewidth]{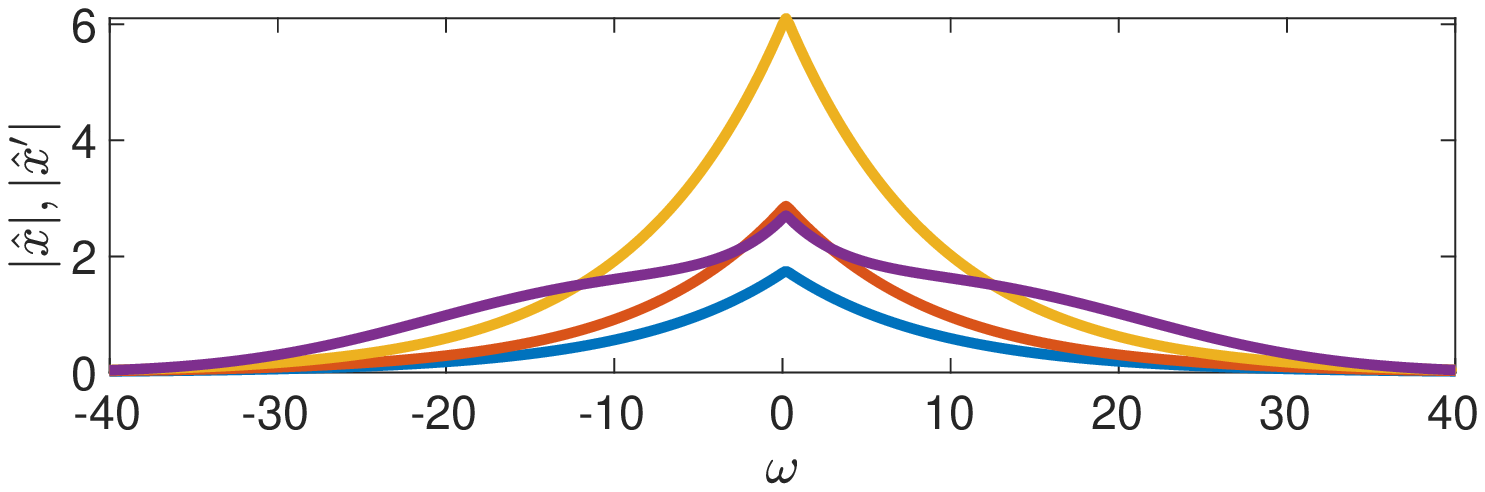}}
			
			\subfloat[Filter ]{\includegraphics[width=0.5\linewidth]{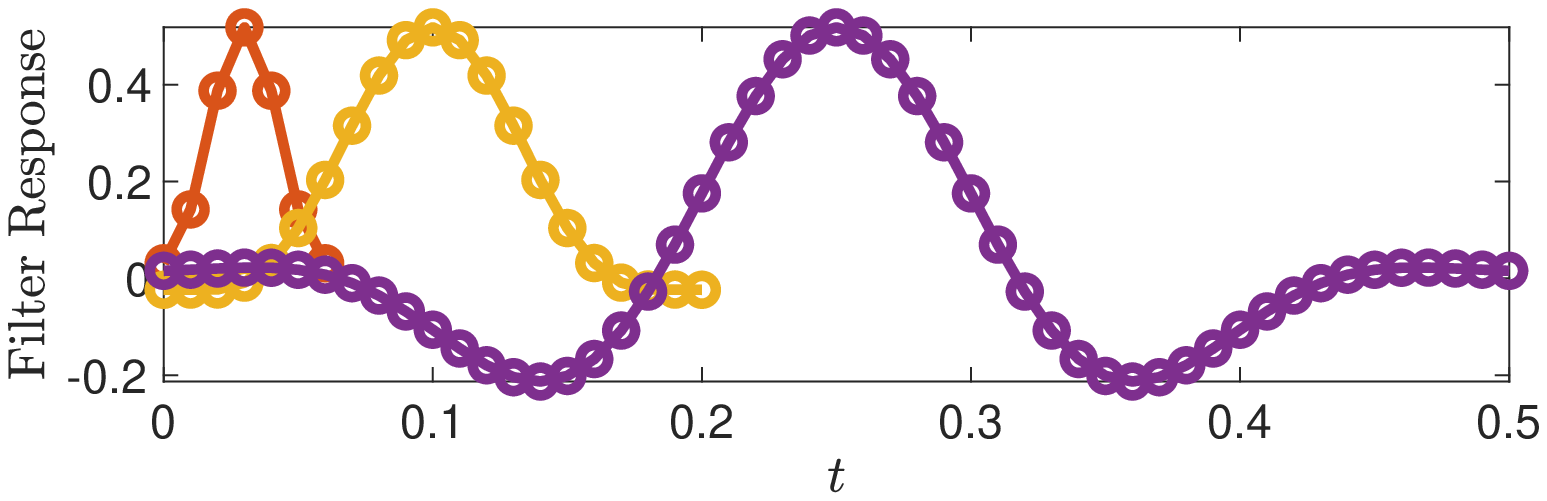}}
			\subfloat[Filter Spectral Response]{\includegraphics[width=0.5\linewidth]{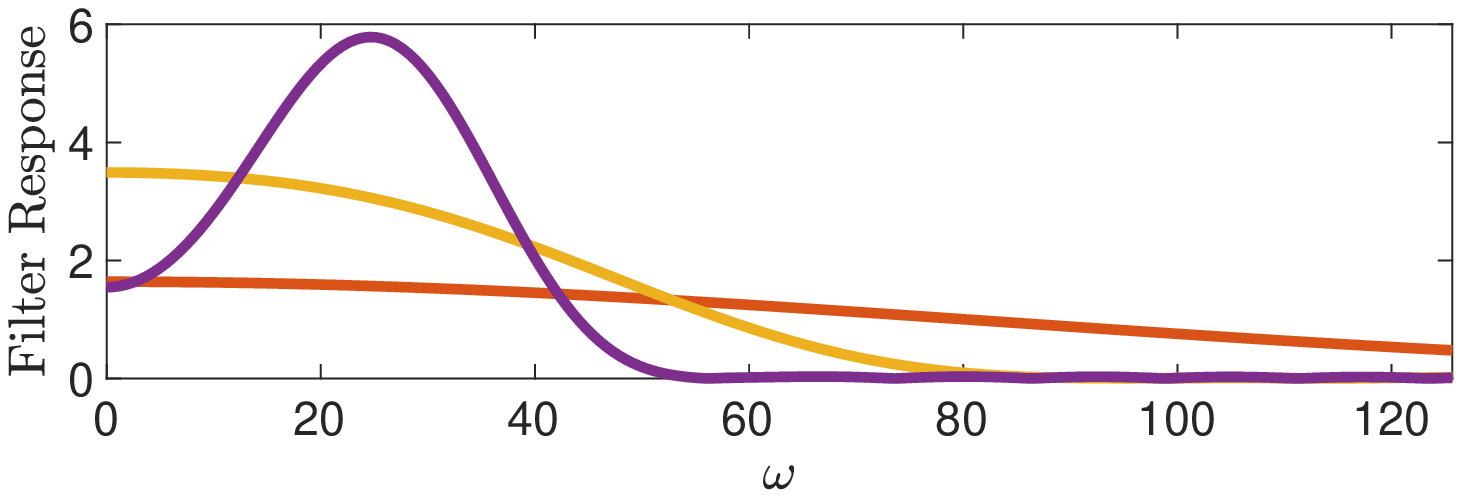}}
			\caption{Same as \ref{fig:filter1}, with a sampling rate 5 times higher.}
			\label{fig:filter2}
		\end{figure}

		Figure \ref{fig:filter2} reproduces the results from figure \ref{fig:filter1} using a sampling rate 5 times higher. It can be seen that filters of much higher order are needed to obtain equivalent results. This is due to the filter frequency resolution, which is proportional to $   n /\Delta t $. Higher filter orders are thus necessary to obtain similar resolutions if higher sampling rates are used.
		


	\section{Interpolation algorithms and their spectral properties} \label{app:interpSpecProp}
	On the TRM methods, described in \S~\ref{sec:transResponse}, simulation checkpoints needs to be saved to disk for later use as an external force. As saving all time steps can require large storage, interpolation between different time steps are typically necessary.
	Three different interpolation methods between flow snapshots are investigated, and named after their smoothness as,
	\begin{itemize}
		\item $ C^0 $: linear interpolation, 
		\item $ C^1 $: cubic interpolation,
		\item $ C^2 $: 5-th order interpolation.
	\end{itemize}
	
	For the $ C^0 $ interpolation method, coefficients of a first order polynomial are chosen as to match the function value at the interval limits. For the $ C^1 $ method, coefficients of a cubic polynomial are chosen to match desired values and first derivatives at the interval limits, with the derivatives  estimated via a second order finite difference scheme. Finally, for the $ C^2 $ method, coefficients of a 5-th order polynomial are chosen  to match values up to the second derivative at the interval limits, with first and second derivatives obtained with a 5-point centred scheme. At the edges, the following are used:
	\begin{itemize}
		\item $ C^1 $ : first order non-centred differentiation is used at the edge points. 
		\item $ C^2 $ : first order non-centred differentiation is used for the first derivative at the edge points, with second derivative set to zero. On the next point, second order differentiation schemes are used to compute the first and second derivative.
	The lower accuracy obtained at the edges has limited impact on the method, as in these regions forcing and responses have vanishingly small amplitudes, so the absolute errors introduced are not relevant.
	\end{itemize}

\begin{figure}
		\centering
		\subfloat[3 poins/cycle]{		\includegraphics[width=.45\linewidth]{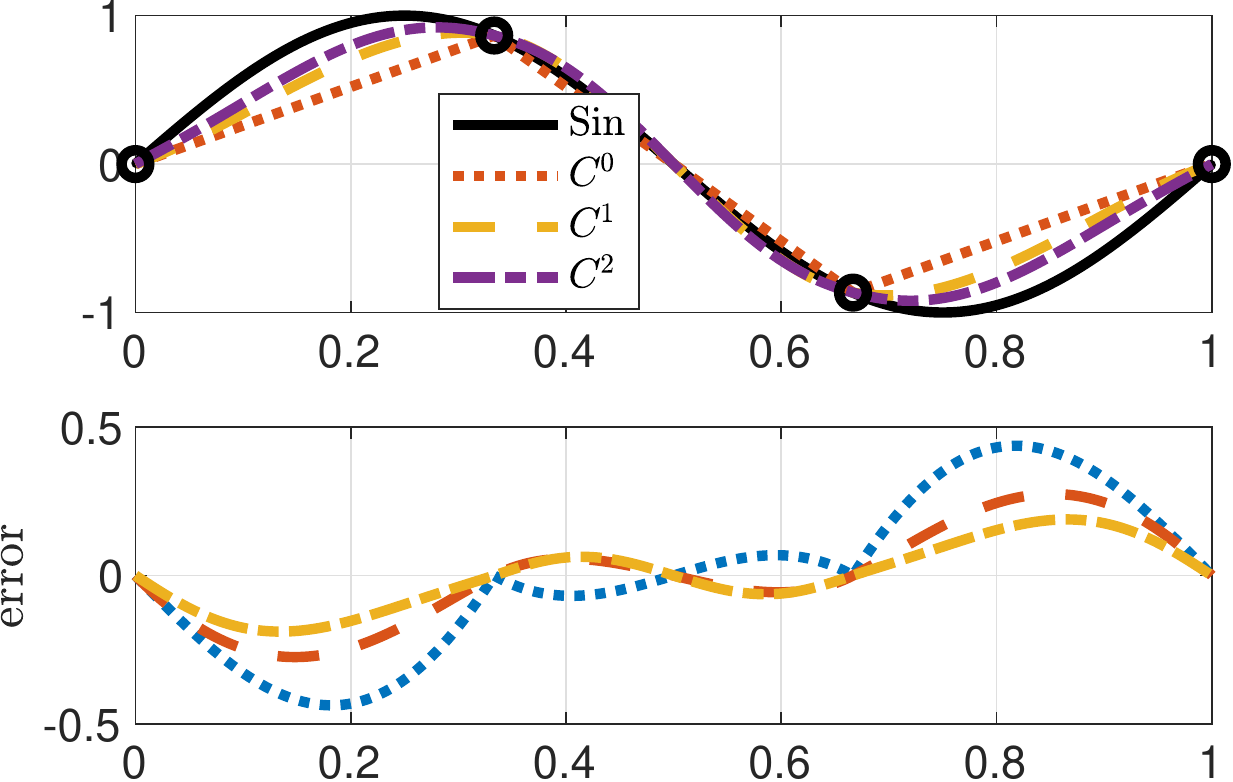} }
		\subfloat[4 poins/cycle]{		\includegraphics[width=.45\linewidth]{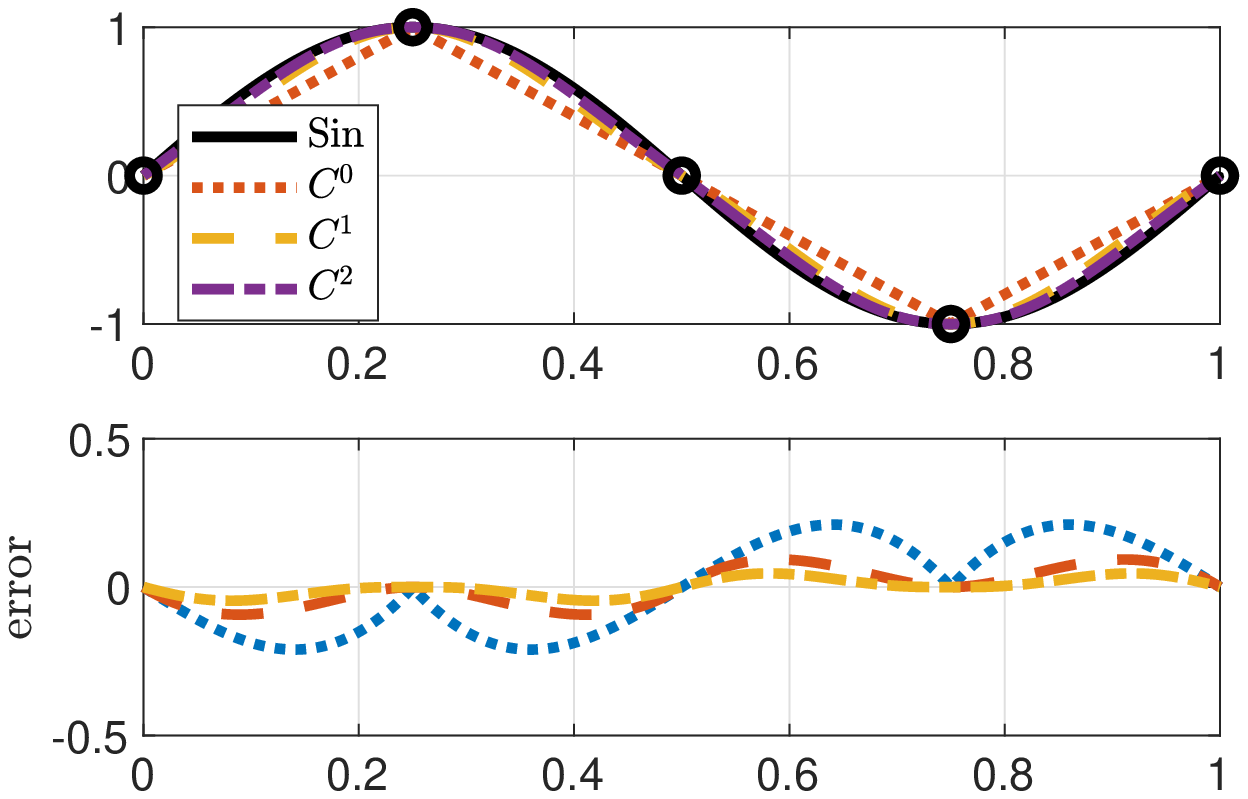} }
		\caption{Interpolation of a sinusoidal signal with the three proposed methods.}
		\label{fig:interpExample}
	\end{figure}

\begin{figure}
		\centering
		\subfloat[Error norm]{		\includegraphics[width=.45\linewidth]{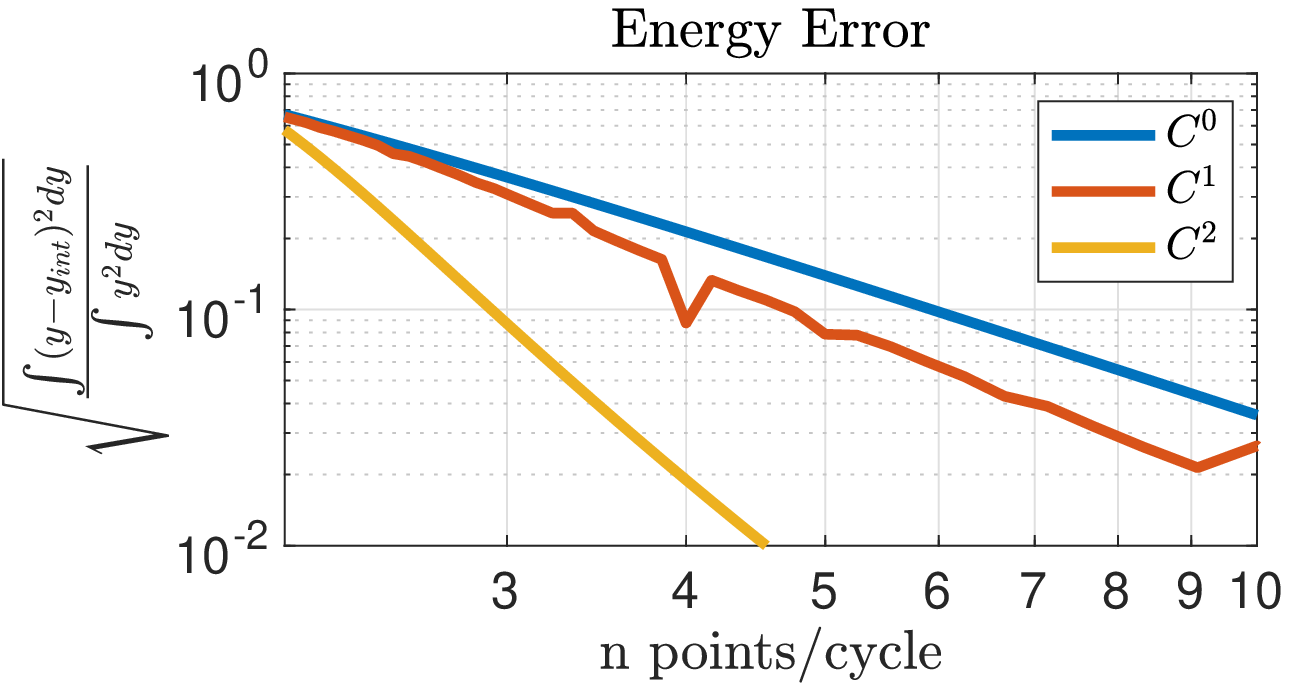} }
		\subfloat[1-st harmonic error]{		\includegraphics[width=.45\linewidth]{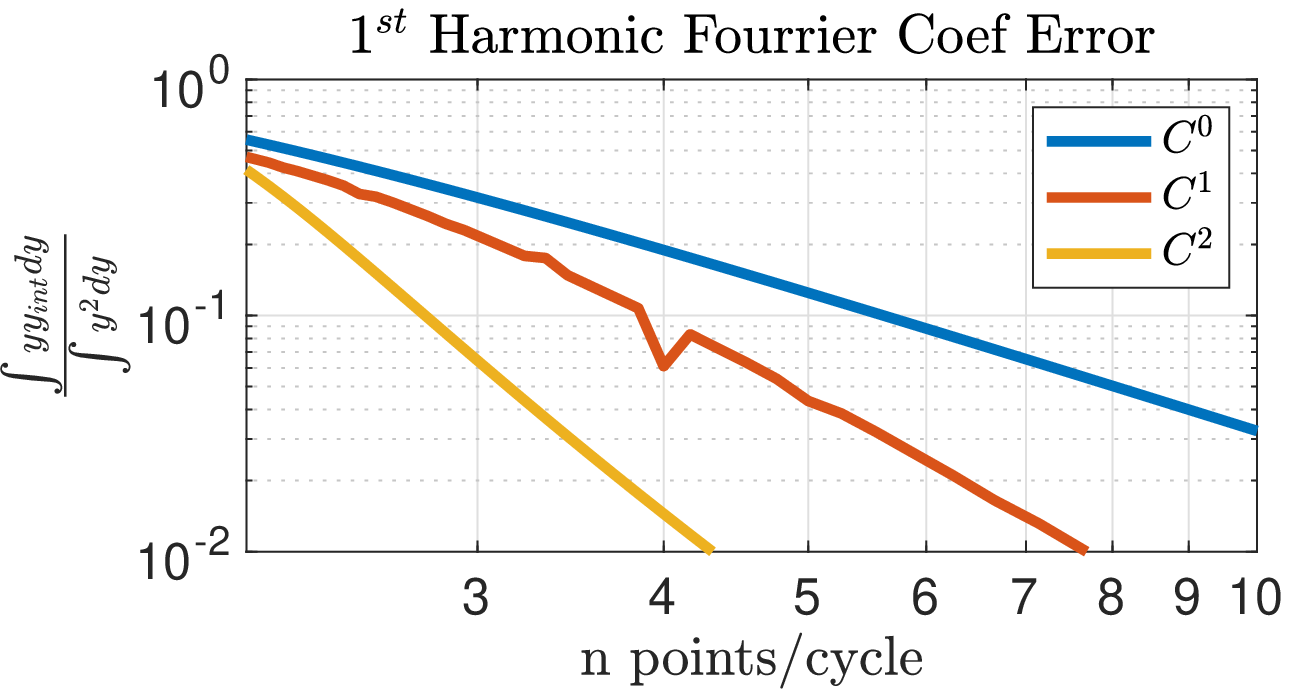} }
		\caption{Interpolation error norm, and error on the 1-st Fourier coefficient.}
		\label{fig:intererror}
	\end{figure}

\begin{figure}
	\centering
	\subfloat[Interpolation Leakage using 2.5 points/cycle]{		\includegraphics[width=.45\linewidth]{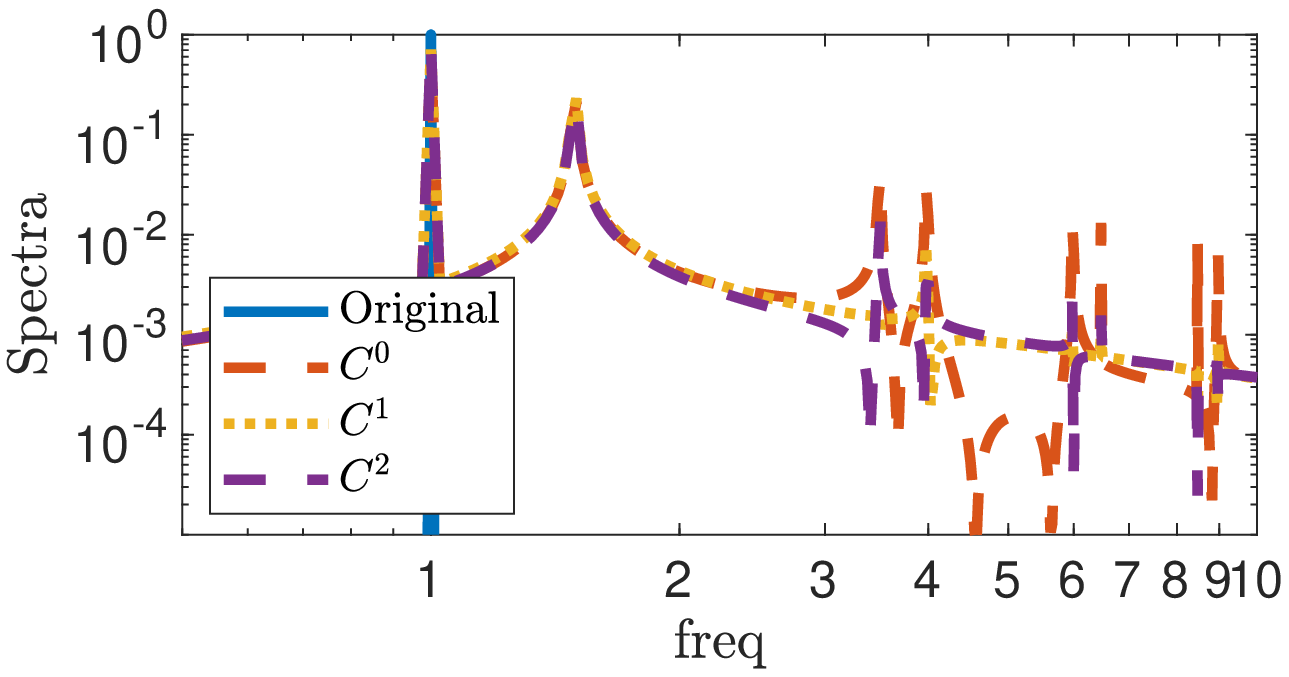} }
		\subfloat[Interpolation Leakage using 4.0 points/cycle]{			\includegraphics[width=.45\linewidth]{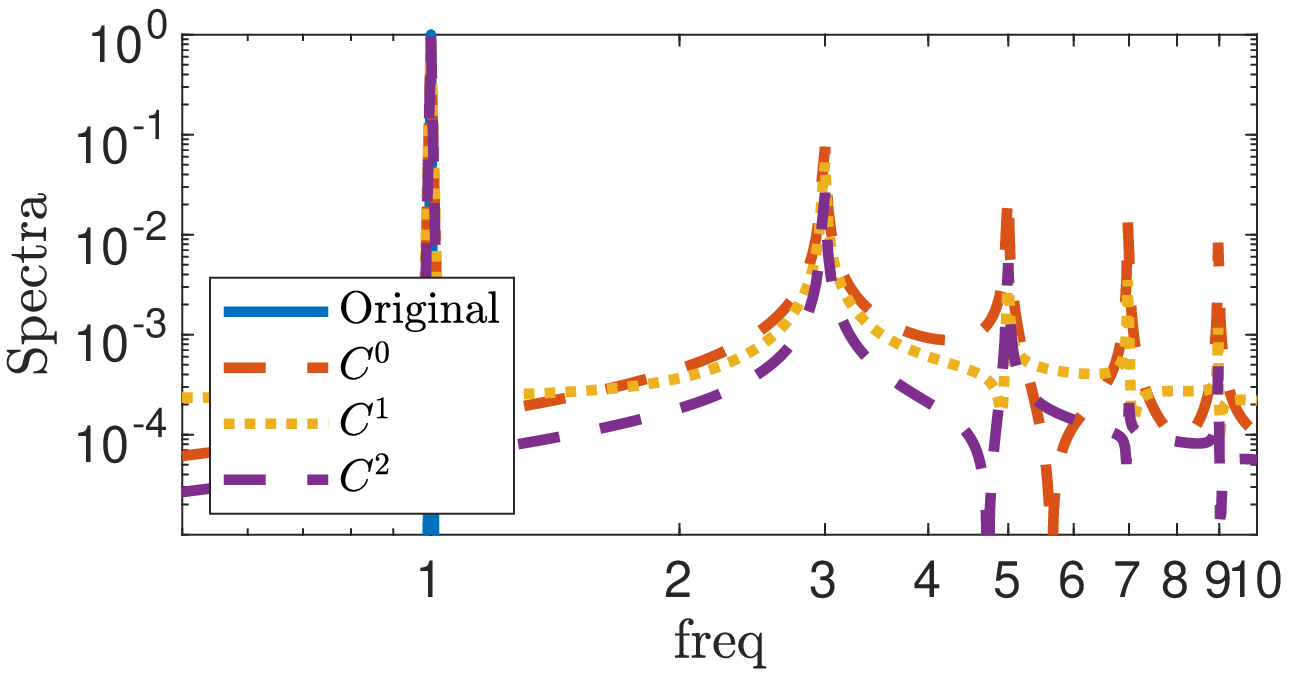} }
	\caption{Spectral content of the interpolated signals.}
	\label{fig:interleak}
	\end{figure}

	Figure \ref{fig:interpExample} shows the performance of each approach for the interpolation of a sinusoidal signal, which reflects errors expected on the Fourier components of the signal. Figure \ref{fig:intererror} presents errors associated with each method, and figure \ref{fig:interleak} shows the spectral content of the interpolated functions. This parameter is important as large errors in the 1-st harmonic create an artificial increase/decrease of resolvent gains. To obtain errors of the order $ 10^{-2} $, approximately 4.5 points/cycle should be used. Interpolated signals with $ n $ points per cycle are seen to generate the first spurious frequency peak at a frequency $ n-1 $ times the frequency of the original signal.
	
			\section{Code Validation} \label{app:validation}
	The implementation of the method on the \emph{Nek5000} code was validated on a channel flow with Reynolds number 50, based on the centreline velocity and the channel half-width. Validations were made for 2D and 3D cases. Domain lengths of $ 2\pi $ in the streamwise direction and $ 0.01 $ in the spanwise directions, for 3D simulations, were used. Periodic conditions in the stream and spanwise directions were used. With such narrow spanwise length, the dominant spanwise wavenumber is $ \beta=0 $ at all frequencies. Results obtained with the implemented code were compared against standard tools based on the decomposition of perturbations in spanwise and streamwise wavenumbers, previously used in \cite{nogueira2020forcing}. Wavenumbers were looped over to search for the dominant gains at each frequency. 
	
	The time marching method closely reproduces the leading gains, validating the implementation.
	
	\begin{figure}
		\centering
		\subfloat[2D]{\includegraphics[width=0.5\linewidth]{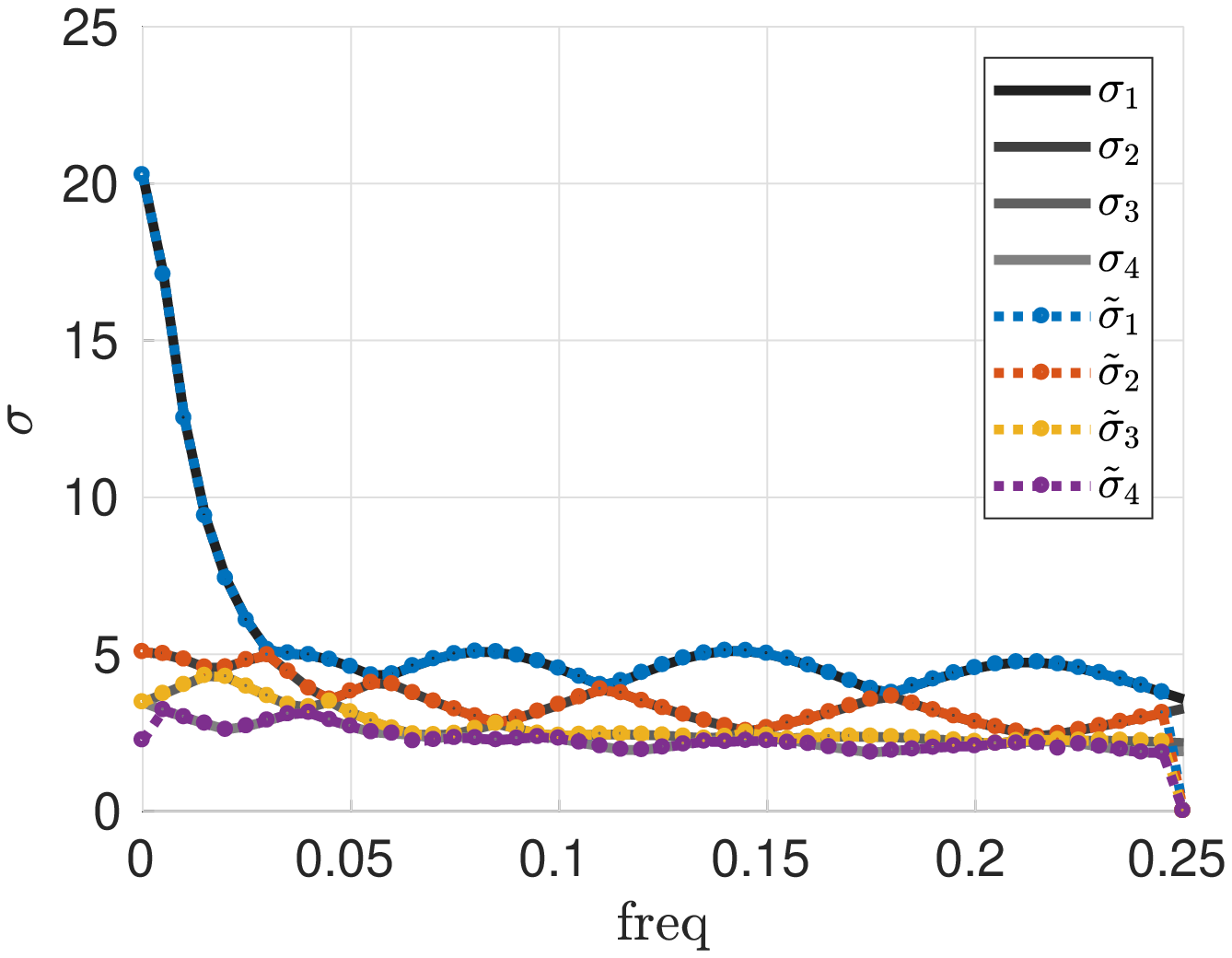}}
		\subfloat[3D]{\includegraphics[width=0.5\linewidth]{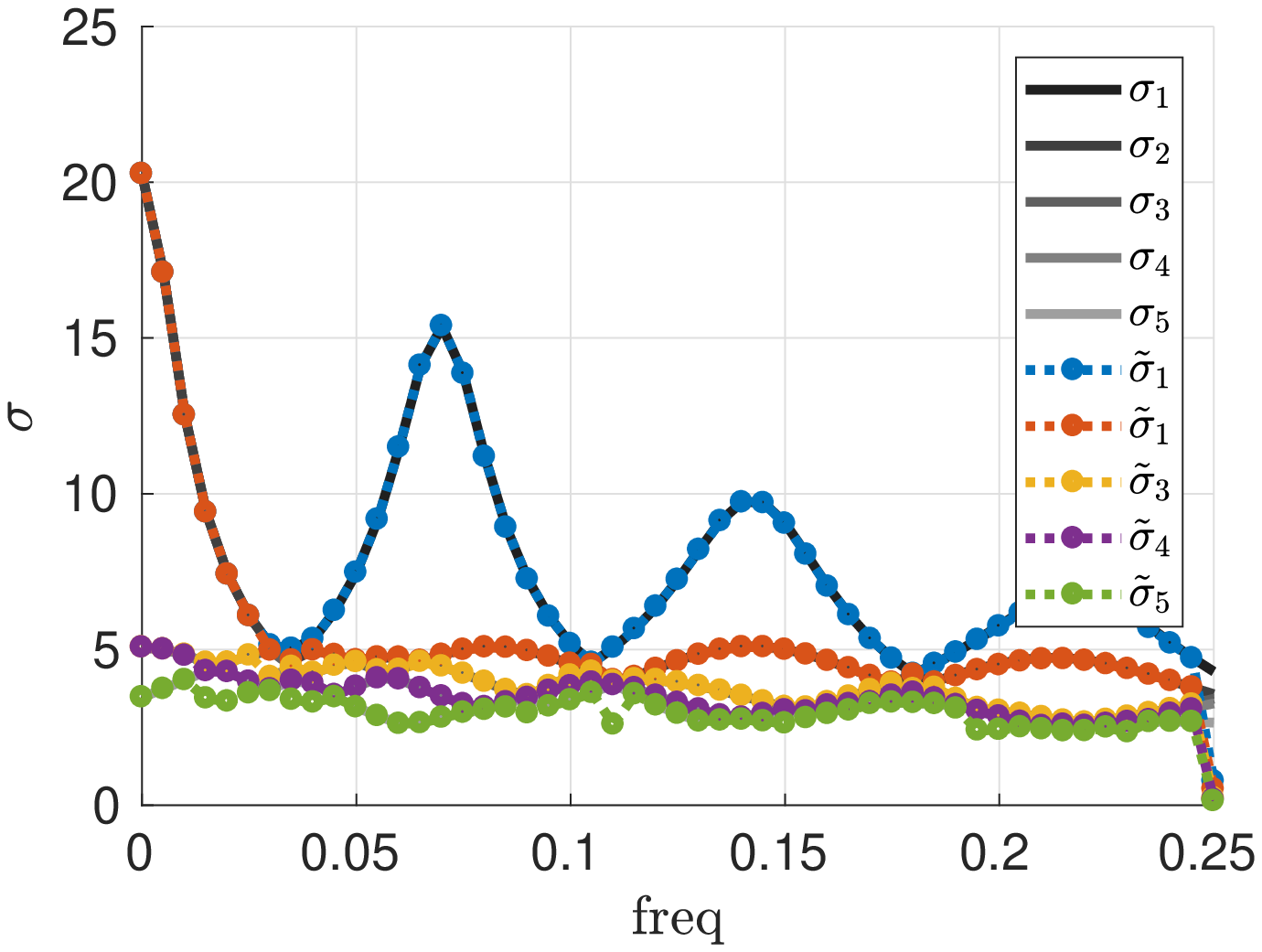}}
		\caption{Code validation on a laminar channel flow.}
		\label{fig:validation}
	\end{figure}

\bibliographystyle{abbrvjfm}


\end{document}